\documentclass[apj, twocolappendix, numberedappendix, appendixfloats]{emulateapj}
\pdfoutput=1 
\usepackage{amsmath,amstext}
\usepackage[T1]{fontenc}
\usepackage{apjfonts}
\usepackage[breaklinks,colorlinks,citecolor=blue]{hyperref} 
\usepackage[figure,figure*]{hypcap}

\newcommand{\Msun}{$M_{\odot}$}

\newcommand{\Vmax}{$V_\mathrm{max}$}

\newcommand{\zminsample}{0.02}
\newcommand{\zmaxsample}{0.06}
\newcommand{\Nes}{$\sim$110'000}

\newcommand{\alphablue}{-1.21}
\newcommand{\logphistarblue}{-2.43}
\newcommand{\logmstarblue}{10.6}
\newcommand{\alphamergers}{$\alpha = -0.55 \pm 0.08$}

\newcommand{\fET}{8.44\%}

\newcommand{\fLT}{33.29\%}
\newcommand{\mergersblue}{55}
\newcommand{\postmergersblue}{87}
\newcommand{\ERDFpaper}{Weigel et al. 2017 in prep.}
\defcitealias{Peng:2010aa}{P10}
\defcitealias{Darg:2010aa}{D10}

\shorttitle{Major merger quenching}
\shortauthors{Weigel et al.}

\begin{document}

\title{Galaxy Zoo: major galaxy mergers are not a significant quenching pathway\footnotemark[*]}

\author{
	Anna K. Weigel \altaffilmark{1},
	Kevin Schawinski \altaffilmark{1},
	Neven Caplar \altaffilmark{1},
	Alfredo Carpineti \altaffilmark{2},
	Ross E. Hart \altaffilmark{3},
	Sugata Kaviraj \altaffilmark{4, 5},
	William C. Keel \altaffilmark{6}, 
	Sandor J. Kruk \altaffilmark{7},
	Chris J. Lintott \altaffilmark{7},
	Robert C. Nichol \altaffilmark{8, 9},
	Brooke D. Simmons \altaffilmark{10} 
	and Rebecca J. Smethurst \altaffilmark{3, 7}
	}
	
\footnotetext[*]{This publication has been made possible by the participation of more than 100 000 volunteers in the Galaxy Zoo project. Their contributions are individually acknowledged at \url{http://authors.galaxyzoo.org}.}
\altaffiltext{1}{Institute for Astronomy, Department of Physics, ETH Zurich, Wolfgang-Pauli-Strasse 27, CH-8093 Zurich, Switzerland}
\altaffiltext{2}{Blackett Laboratory, Imperial College London, London SW7 2AZ, UK}
\altaffiltext{3}{School of Physics and Astronomy, The University of Nottingham, University Park, Nottingham NG7 2RD, UK}
\altaffiltext{4}{Centre for Astrophysics Research, University of Hertfordshire, College Lane, Hatfield AL10 9 AB, UK}
\altaffiltext{5}{Worcester College, Oxford, UK}
\altaffiltext{6}{Department of Physics and Astronomy, University of Alabama, Box 879324, Tuscaloosa, AL 35487, USA}
\altaffiltext{7}{Oxford Astrophysics, Denys Wilkinson Building, Keble Road, Oxford OX1 3RH, UK}
\altaffiltext{8}{Institute of Cosmology \& Gravitation, University of Portsmouth, Dennis Sciama Building, Portsmouth PO1 3FX, UK}
\altaffiltext{9}{South East Physics Network, www.sepnet.ac.uk}
\altaffiltext{10}{Center for Astrophysics and Space Sciences (CASS), Department of Physics, University of California, San Diego, CA 92093, USA}

\begin{abstract}
We use stellar mass functions to study the properties and the significance of quenching through major galaxy mergers. In addition to SDSS DR7 and Galaxy Zoo 1 data, we use samples of visually selected major galaxy mergers and post merger galaxies. We determine the stellar mass functions of the stages that we would expect major merger quenched galaxies to pass through on their way from the blue cloud to the red sequence: 1: major merger, 2: post merger, 3: blue early type, 4: green early type and 5: red early type. Based on the similar mass function shapes we conclude that major mergers are likely to form an evolutionary sequence from star formation to quiescence via quenching. Relative to all blue galaxies, the major merger fraction increases as a function of stellar mass. Major merger quenching is inconsistent with the mass and environment quenching model. At $z\sim0$ major merger quenched galaxies are unlikely to constitute the majority of galaxies that transition the green valley. Furthermore, between $z\sim 0 - 0.5$ major merger quenched galaxies account for $1 - 5 \%$ of all quenched galaxies at a given stellar mass. Major galaxy mergers are therefore not a significant quenching pathway, neither at $z\sim0$ nor within the last 5 Gyr. The majority of red galaxies must have been quenched through an alternative quenching mechanism which causes a slow blue to red evolution.
\end{abstract}


\section{Introduction}
The physical cause of the cessation of star formation is an open question in astrophysics today. In the local Universe, galaxies fall into two broad categories: spiral or late type galaxies, which mostly have a blue optical colour, and elliptical or early type galaxies, which are primarily optically red (but also see: \citealt{Schawinski:2009aa, Masters:2010aa}). In the colour-mass and colour-magnitude diagrams galaxies separate into the `blue cloud' and the `red sequence' \citep{Bell:2003aa, Baldry:2004aa, Martin:2007aa, Taylor:2015aa}. Between the blue cloud and the red sequence lies the so-called `green valley' \citep{Bell:2003aa, Martin:2007aa, Faber:2007aa, Fang:2012aa, Schawinski:2014aa}, a transition zone that contains both late and early type galaxies. Blue cloud galaxies are also often referred to as galaxies on the `main sequence' \citep{Brinchmann:2004aa, Salim:2007aa, Noeske:2007aa, Daddi:2007aa, Elbaz:2007aa, Peng:2010aa, Lilly:2013aa, Speagle:2014aa, Lee:2015aa, Tomczak:2016aa, Kurczynski:2016aa}. In star formation rate (SFR) versus stellar mass space they lie on an almost linear relation. Red early type galaxies are quiescent. They have significantly lower SFRs than blue late types and thus lie below the main sequence.  The bimodality in colour-mass and colour-magnitude space and the existence of the main sequence imply that blue galaxies are likely to shut down their star formation at some point during their lifetime. A significant decrease in the SFR causes them to transition from the blue cloud to the green valley and finally to the red sequence. In combination with a morphological transformation from spiral to elliptical, this evolution could explain the existence of the red sequence. We refer to the physical process that causes blue galaxies to shut down their star formation as quenching. 

A variety of physical processes that could cause star formation quenching have been proposed. These can be classified into internal and external processes. Internal processes include AGN feedback \citep{Silk:1998aa, Di-Matteo:2005aa,Schawinski:2006aa, Kauffmann:2007aa, Georgakakis:2008aa, Hickox:2009aa, Cattaneo:2009ab, Fabian:2012aa, Bongiorno:2016aa, Kaviraj:2017aa} and secular processes \citep{Kormendy:2004aa, Masters:2011aa, Cheung:2013aa}. Externally, quenching could be correlated with the environment \citep{Gunn:1972aa, Larson:1980aa, Balogh:2000aa, Woo:2015aa, Knobel:2015aa, Peng:2015aa} or with the occurrence of major galaxy mergers \citep{Sanders:1988aa, Mihos:1996aa, Springel:2005ac, Di-Matteo:2005aa,  Hopkins:2006aa, Croton:2006aa,Hopkins:2008ab, Khalatyan:2008aa, Somerville:2008aa}. Of course we could also imagine a combination of different processes. For example, (\citealt{Peng:2010aa} hereafter \citetalias{Peng:2010aa} and \citealt{Peng:2012aa}) use external (`environment quenching') and  mass dependent, likely internal (`mass quenching')  processes to reproduce the stellar mass function of red galaxies with their phenomenological model. 

Our aim is to study the classical quenching model based on major mergers. Mergers between gas rich galaxies of comparable mass cause most of the galaxies' gas to be driven to the new center. This can ignite both a starburst and an AGN. Star formation and AGN feedback can expel the gas from the galaxy, preventing further star formation. The now elliptical galaxy leaves the blue cloud and  crosses the green valley before settling on the red sequence \citep{Sanders:1988aa, Mihos:1996aa, Springel:2005ac, Di-Matteo:2005aa,  Hopkins:2006aa, Croton:2006aa,Hopkins:2008ab, Khalatyan:2008aa, Somerville:2008aa}. 

We use stellar mass functions of galaxies that are transitioning from the blue cloud to the red sequence to study the properties and the significance of this quenching process. Stellar mass functions are an important statistical measure that allow us to study and infer the properties of a large sample of galaxies. Specifically, they allow us to probe if quenching through major mergers includes a mass dependence. Furthermore, we can constrain the relative amount of time spent in stages along the sequence and measure the merger fraction. By comparing the stellar mass function shapes of major mergers and red galaxies we can also test if merger quenching can account for all quenched galaxies or if an additional quenching channel is necessary. 

To construct these stellar mass functions, we rely on morphological classifications from Galaxy Zoo\footnote{\url{http://www.galaxyzoo.org}}. Besides the classifications from Galaxy Zoo 1 (GZ1; \citealt{Lintott:2008aa,Lintott:2011aa}), we also use the major merger sample by  \citetalias{Darg:2010aa} (hereafter \citetalias{Darg:2010aa}) and \cite{Darg:2010ab} and the post merger sample by \cite{Carpineti:2012aa}. Our analysis is thus based on the visual classifications of over 100 000 Galaxy Zoo volunteers.  

We determine the stellar mass functions of galaxies along the major merger quenching sequence by using the method introduced in \cite{Weigel:2016aa}. This approach is based on the combination of three independent methods ($1/V_{\rm max}$: \citealt{Schmidt:1968aa}, STY: \citealt{Sandage:1979aa}, SWML: \citealt{Efstathiou:1988aa}). Blue, star forming and red, quiescent galaxies are usually well fit by single and double Schechter functions, respectively (e.g. \citealt{Li:2009aa}; \citetalias{Peng:2010aa}; \citealt{Pozzetti:2010aa, Baldry:2012aa, Ilbert:2013aa, Muzzin:2013aa}). Yet it is important to note that when fitting the stellar mass functions, we are not making any a priori assumptions on which galaxy subsample should be fit with a single and a double Schechter function. We use a likelihood ratio test to determine the better fitting model. 

This paper is organized as follows. In Section \ref{sec:data} we introduce the galaxy, the major merger and the post merger sample and give a brief overview of the stellar mass function method used in  \cite{Weigel:2016aa}. Section \ref{sec:method} represents the first part of the paper and is purely data driven: we introduce the stellar mass functions of major mergers and post mergers, determine the merger fraction and test if our measurements are consistent with the phenomenological model by \citetalias{Peng:2010aa}. In the second part of the paper we use these stellar mass functions to investigate the process of major merger quenching. First, we introduce and motivate our assumptions in Section \ref{sec:method}. Second, in Section \ref{sec:analysis} we apply these assumptions to our measurements.   
This is followed by a discussion and a summary in Sections \ref{sec:discussion} and \ref{sec:summary}, respectively. 

Throughout this paper we assume a $\Lambda$CDM cosmology with $h_0 = 0.7$, $\Omega_{\rm m} = 0.3$ and $\Omega_\Lambda = 0.7$ \citep{Komatsu:2011aa}. 

\section{Data}\label{sec:data}
\begin{figure*}
	\begin{center}
	\includegraphics[width=0.66\textwidth]{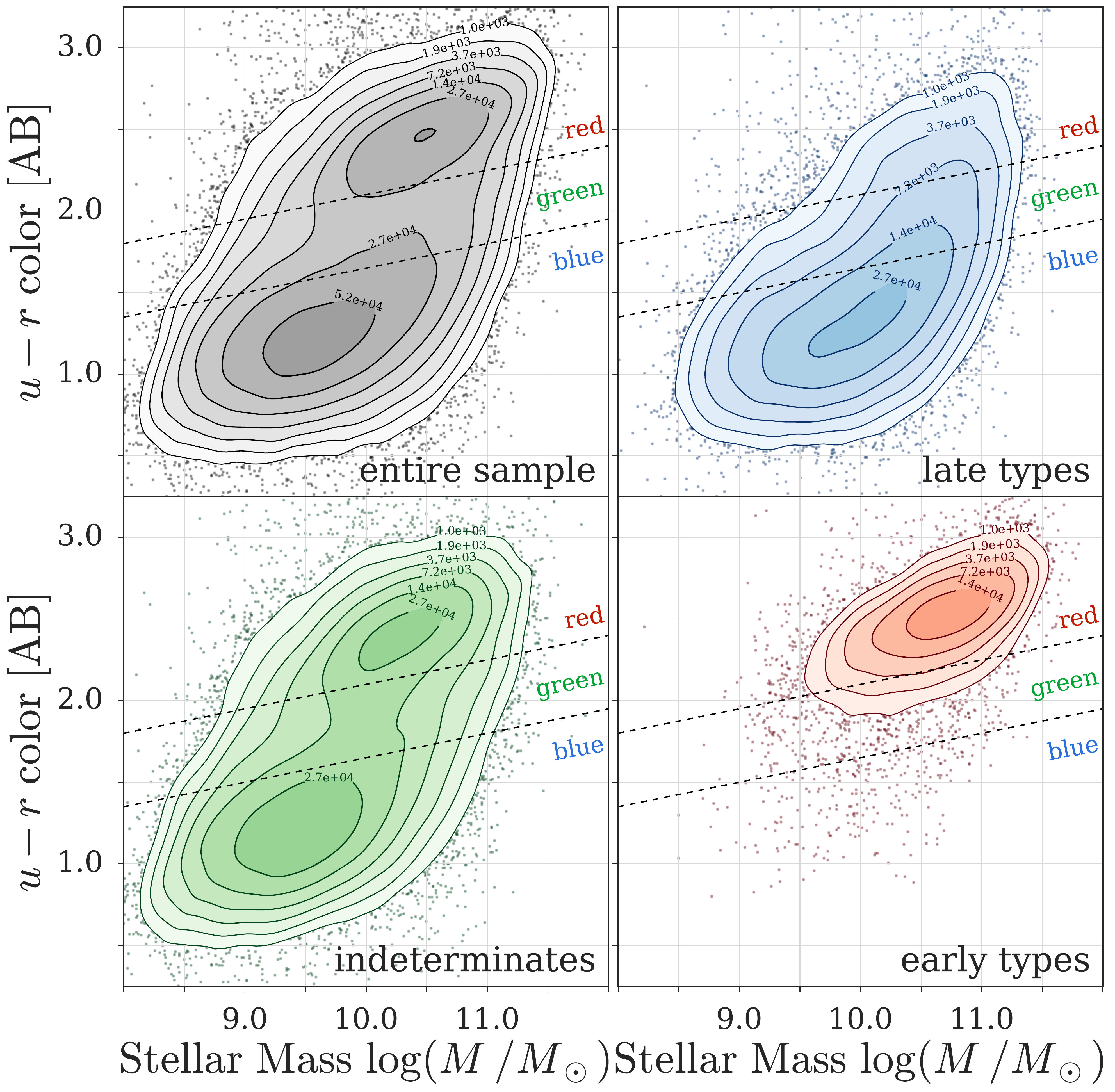}
	\caption{\label{fig:cm} Colour mass diagrams for the entire galaxy sample, indeterminates and late and early types. All colours are dust and k- corrected. The dashed lines indicate our definition of the green valley (equations \ref{eq:cut_red}, \ref{eq:cut_blue}). Out of the \Nes\ objects in the entire galaxy sample, \fLT\ and \fET\ are classified as being late and early type galaxies, respectively. For the remaining galaxies the probability of being a late or an early type galaxy lies below the vote fraction threshold. They are categorized as indeterminate. This figure illustrates that splitting the sample by colour is not equivalent to splitting the sample by morphology since not all late type galaxies are blue \protect\citep{Masters:2010aa} and not all early type galaxies are red \protect\citep{Schawinski:2009aa}. The contours represent equal steps in in log space and show the number of objects.}
	\end{center}
\end{figure*}

\subsection{The SDSS galaxy sample}\label{sec:galaxy_sample}
For our analysis we use data from the seventh data release (DR7) of the Sloan Digital Sky Survey (SDSS; \citealt{York:2000aa, Abazajian:2009aa}). We extract spectroscopic redshift and magnitude values from the New York Value-Added Galaxy Catalog (NYU VAGC; \citealt{Blanton:2005aa, Padmanabhan:2008aa}). For the stellar masses measurements we use \cite{Brinchmann:2004aa} recorded in the Max Planck Institute for Astrophysics \ John Hopkins University (MPA JHU; \citealt{Kauffmann:2003ab, Brinchmann:2004aa, Salim:2007aa}) catalog. These stellar mass estimates are based on fits to the photometry and model spectra by \cite{Bruzual:2003aa} and are in good agreement to the $4000$ \AA\ and H$\delta_{\rm A}$ based measurements by \cite{Kauffmann:2003ab}\footnote{\url{http://wwwmpa.mpa-garching.mpg.de/SDSS/DR7/mass_comp.html}}. 

We cross match to the GZ1 catalog\footnote{\url{http://data.galaxyzoo.org}} (table 2 in \citealt{Lintott:2011aa}, also see \citealt{Lintott:2008aa, Land:2008aa}) to obtain morphological classifications for all galaxies. Each object in the sample was classified $38$ times on average by over $100\ 000$ Galaxy Zoo volunteers. GZ1\footnote{\url{http://zoo1.galaxyzoo.org/}} users were given six possible classifications for each galaxy (`Elliptical galaxy', `Clockwise/Z-wise spiral galaxy', `Anti-clockwise/S-wise spiral galaxy', `Spiral galaxy other (e.g. edge on)', `Star or do not know', `Merger'). These can be summarized into `elliptical' (E) and `combined spiral' (CS = `Clockwise/Z-wise spiral galaxy' + `Anti-clockwise/S-wise spiral galaxy' + `Spiral galaxy other (e.g. edge on)') galaxies. The likelihood of a galaxy having a spiral or an elliptical morphology depends on the fraction of users that have classified the galaxy as such. This is referred to as the vote fraction.

We base our analysis on the `elliptical', `spiral' and `uncertain' type flags. These flags are based on vote distributions that have been corrected for classification bias (\citealt{Bamford:2009aa}, see \citealt{Willett:2013aa} and \citealt{Hart:2016aa} for Galaxy Zoo 2). High redshift galaxies are more likely to be categorized as ellipticals since they appear fainter and smaller which makes it more difficult for the classifier to recognize morphological features. To correct for this effect elliptical and combined spiral galaxies with raw vote fractions above $80\%$ are chosen to compute the elliptical-to-spiral ratio. The raw vote distributions are then debiased by assuming that there is no redshift evolution in this morphological ratio within bins of luminosity and size. Galaxies with debiased vote fractions above $80\%$ in the elliptical and spiral categories are then flagged as `elliptical' and `spiral', respectively. Galaxies for which the debiased vote fractions in both the elliptical and the combined spiral category lie below $80\%$ are flagged as `uncertain'. We note that the GZ1 interface did not allow users to classify galaxies as `uncertain'. The `uncertain' flag simply reflects the fact that a galaxy's spiral and elliptical probabilities lie below the corresponding thresholds. \cite{Schawinski:2014aa} argue that the majority of galaxies in the `uncertain' category show late type characteristics, whereas only a small fraction might be misclassified early types. By using a high debiased vote fraction cut of $80\%$ we eliminate some of the nuances in galaxy morphologies. Yet the resulting clean early type sample allows the inference of a broad picture of galaxy evolution. We refer to galaxies which are flagged as `elliptical', `spiral' and `uncertain' as early types, late types and indeterminates.

To be able to correct for dust, we use the absorption- and emission-line measurements from OSSY \citep{Oh:2011aa}. As an environment estimate, we include the overdensity measurements from \cite{Weigel:2016aa} which are based on a 5th nearest neighbor approach ($M > 10^9$ \Msun, recession velocity range $\pm 1000 \mathrm{km}/\mathrm{s}$). We also add halo mass measurements, spectral completeness values and the classification into centrals and satellites from the \cite{Yang:2007aa} catalog.  We limit our main sample to the redshift range between \zminsample\ and \zmaxsample\ and to objects of the MPA JHU spectral type `GALAXY'.  We refer to the sample of galaxies that lie within this redshift range, have the correct spectroscopic classification and for which stellar masses, morphological classifications and environment and emission line measurements are available as the `entire galaxy sample'. For more details on this sample and the overdensity measurement see \cite{Weigel:2016aa}. 

\subsection{Colour cuts}\label{sec:colour}
We use the colour-mass diagram \citep{Bell:2003aa, Baldry:2004aa, Martin:2007aa, Faber:2007aa, Schawinski:2014aa} to split our sample into red, green and blue galaxies. We use the Petrosian flux values from the NYU VAGC and apply a dust and k-correction. We k-correct to redshift zero using the \textsc{kcorrect} \textsc{idl} package (version 4.2) by \cite{Blanton:2007aa} and use the Calzetti law \citep{Calzetti:2000aa} with $E(B-V)$ values from OSSY ([EBV\_STAR], \citealt{Oh:2011aa}) to correct for internal dust extinction. We show the colour-mass diagram for our main sample in Fig. \ref{fig:cm}. 

We use the colour definitions from \cite{Weigel:2016aa} and refer to sources lying above :
\begin{equation}\label{eq:cut_red}
u - r\ (\log M) = 0.6 + 0.15 \times \log M
\end{equation}
as being red and to galaxies below 
\begin{equation}\label{eq:cut_blue}
u- r\ (\log M) = 0.15 + 0.15 \times \log M
\end{equation}
as being blue. Objects between equations \ref{eq:cut_red} and \ref{eq:cut_blue} are part of the green valley \citep{Bell:2004aa, Martin:2007aa, Fang:2012aa, Schawinski:2014aa} and are referred to as being green.

The colour-mass diagram in Fig. \ref{fig:cm} illustrates that splitting  the sample by colour and morphology yields different results. Not all early types are red \citep{Schawinski:2009aa} and not all late types are blue \citep{Masters:2010aa}. 

\subsection{The major merger sample}\label{sec:merger_sample}
We use the Galaxy Zoo merger sample by \citetalias{Darg:2010aa} and \cite{Darg:2010ab}. The sample is based on SDSS DR6 and contains 3003 visually classified merging systems in the redshift range  $0.005- 0.1$.

\citetalias{Darg:2010aa} based their sample on the GZ1 morphological classifications (see Sect. \ref{sec:galaxy_sample}). For each source, \citetalias{Darg:2010aa} calculated the ratio of the number of people who classified this objects as a merger to the total number of classifications of this source. The weighted-merger-vote fraction $f_{\rm m}$ of this object is then defined as this ratio multiplied by a weighting factor that represents the reliability of all users that classified the object. If $f_{\rm m}$ is equal to its minimum value $0$, the galaxy is unlikely to be a merger. If $f_m$ is equal to its maximum value 1 the galaxy has consistently been classified as a merger. In their catalog \citetalias{Darg:2010aa} only include galaxies for which $f_{\rm m} > 0.4$. 

\citetalias{Darg:2010aa} determine stellar masses for all galaxies in their sample by fitting two-component star formation histories to the photometry. These fits are based on \cite{Maraston:1998aa,Maraston:2005aa} stellar models, a Salpeter \citep{Salpeter:1955aa} initial mass function, stellar populations with fixed solar metallicity and variable ages and a dust implementation according to the Calzetti law \citep{Calzetti:2000aa}. SDSS spectra and thus MPA JHU \citep{Kauffmann:2003ab, Brinchmann:2004aa, Salim:2007aa} stellar mass measurements for both galaxies involved in the merger are only available for $23\%$ of all merging systems in the \citetalias{Darg:2010aa} catalog. We thus use the photometry based stellar mass measurements by \citetalias{Darg:2010aa} to restrict ourselves to mergers between galaxies with a mass ratio within $1/3 < M_1/M_2 < 3$.

While we use the stellar mass estimates by \citetalias{Darg:2010aa} to select major mergers, we use the stellar mass values by \cite{Brinchmann:2004aa} for the construction of the major merger mass function to ensure consistency with the other mass functions presented here. If spectra are available for both merging galaxies we consider the mass of the more massive merging partner in the construction of the major merger mass function. If only one of the two sources has been observed spectroscopically, we take the mass of the source with an available spectrum into account. According ot the \citetalias{Darg:2010aa} mass measurements, this corresponds to the more massive merging partner in $69\%$ of all merging systems with one spectrum. We discuss the effect of this approach on the merger mass function shape in Section \ref{sec:merger_smf_mass}.

\citetalias{Darg:2010aa} find that their sample of merging galaxies contains three times as many spiral as elliptical galaxies and is dominated by mergers between spiral galaxies. In the general galaxy population the ratio of spirals to ellipticals is 3:2 for the same redshift range and above the same magnitude limit. \cite{Willett:2015aa} show that on average the \citetalias{Darg:2010aa} merging galaxies pairs lie $\sim 0.3$ dex above the main sequence relation. Using the colour definitions introduced in Section \ref{sec:colour}, we classify $\sim\mergersblue \%$ of all major merger galaxies in our sample as blue. We thus conclude that the majority of merging galaxies in the \citetalias{Darg:2010aa} sample are blue, star forming spirals. 

Galaxy pairs which seem to be merging due to projection effects can easily be eliminated if spectroscopic redshifts are available for both galaxies. When only one of the merging galaxies has an available spectrum, the major merger candidate has to be visually examined for galaxy interactions. \citetalias{Darg:2010aa} argue that galaxies with $f_{\rm m} > 0.4$ are predominately clear major mergers and that decisions regarding possible projection effects only have to be made in rare cases. Stellar projections are excluded based on the SDSS \textsc{PHOTOTAG} `type' classification which indicates whether the possible merging partner is point-like or extended. 

\subsection{The post merger sample}
In the construction of their major merger sample \citetalias{Darg:2010aa} flag objects that were classified as being a major merger by the Galaxy Zoo users and only show a single core. These objects are identified as a single source by SDSS, but show strong perturbations in the outskirts. They are therefore likely to be objects in the late stages of a merger. \citetalias{Darg:2010aa} flag these objects which can no longer be resolved by the SDSS pipeline as `post mergers'. Strong perturbations in the periphery can also be caused by a close encounter with a second galaxy that is no longer in the field of view. These objects are flagged as `fly-bys'. While \citetalias{Darg:2010aa} do not include these perturbed systems in their major merger catalog, \cite{Carpineti:2012aa} investigate the colour and AGN activity of as `post merger' flagged sources relative to early type galaxies.  We refer to galaxies in the \cite{Carpineti:2012aa} sample as post mergers.  

\cite{Carpineti:2012aa} select a sample of spheroidal post mergers and argue that $\sim 55\%$ of sources in this sample are remnants of merging systems which involved at least one late type galaxy. According to our colour definitions, $\sim\postmergersblue\%$ are defined as blue.

We note that the major merger and post merger sample are based on SDSS DR6 while the rest of our analysis is based on SDSS DR7. The galaxies of the SDSS DR6 spectroscopic sample make up $\sim 85\%$ of the galaxies in the SDSS DR7 spectroscopic sample \footnote{See \url{http://classic.sdss.org/dr6/} and \url{http://classic.sdss.org/dr7/}}. Assuming there is no bias in the way the additional galaxies in DR7 were selected, we would expect to find $\sim15\%$ more major merger and post merger galaxies if we select them in the same way from DR7 instead of DR6. For the major merger and post merger mass functions which we will determine below this would result in a constant increase in the normalization by $\sim0.06$ dex. Using SDSS DR6 instead of SDSS DR7 data for the major merger and post merger sample does hence not significantly affect our results.
\subsection{Stellar mass function construction}
To construct stellar mass functions we follow \cite{Weigel:2016aa} and combine the classical 1/\Vmax\ approach developed by \cite{Schmidt:1968aa} with the parametric maximum likelihood method by \cite{Sandage:1979aa} (STY) and the non-parametric step-wise maximum likelihood method (SWML) which was established by \cite{Efstathiou:1988aa} .

In STY, we are assuming that the stellar mass function can be modelled by either a single or a double Schechter function  \citep{Schechter:1976aa}. We estimate the likelihood of both functional forms and use a likelihood ratio test to determine which model provides a better description of the data.

In the figures below we show the 1/\Vmax\ and SWML results with open and filled symbols, respectively. Upper limits according to the two methods are shown with arrows of the same style. The best-fitting Schechter functions according to the STY method are illustrated with solid lines. The corresponding $1\sigma$ errors are shown as shaded regions. 

We define the single Schechter function as: 

\begin{equation} \label{eq:single_Schechter}
\Phi\ d\log M = \ln(10) \Phi^{*} e^{-M/M^{*}} \left(\frac{M}{M^{*}}\right)^{\alpha + 1} d\log M
\end{equation}

and use the following definition for the double Schechter function:

\begin{equation} \label{eq:double_Schechter}
\begin{aligned}
\Phi\ d\log M = & \ln(10) e^{-M/M^{*}} \\
& \times\left[ \Phi^{*}_1 \left(\frac{M}{M^{*}}\right)^{\alpha_1 + 1} +  \Phi^{*}_2 \left(\frac{M}{M^{*}}\right)^{\alpha_2 + 1}\right]\ d\log M.
\end{aligned}
\end{equation}
Note that the $\ln 10$ factor and the $+1$ in the exponent of $M/M^{*}$ is due to the conversion from $dM$ to $d\log M$.

For each sample, we determine the stellar mass completeness as a function of redshift using the technique introduced by \cite{Pozzetti:2010aa}. This approach is based on keeping the mass-to-light ratio of each individual source constant and determining the stellar mass that this object would have if its redshift stayed constant, but its flux was equal to the magnitude limit.  

For the 1/\Vmax\ approach we determine the stellar mass completeness of each subsample and subsequently the \Vmax\ values of each source by using the approach by \cite{Pozzetti:2010aa}. To be able to apply the STY and the SWML method, we also estimate the minimum stellar mass at which each galaxy would still be part of the sample. We do so by keeping the mass-to-light ratio constant and scaling the flux down to the $r$-band flux limit. 

\cite{Weigel:2016aa} show that the three independent mass function estimators deviate at the low mass end. Compared to STY and SWML, the 1/\Vmax\ technique tends to overestimate $\Phi$ (also see \citealt{Efstathiou:1988aa} and \citealt{Willmer:1997aa}) and depends strongly on the shape of the stellar mass completeness function. While STY and SWML might be less commonly used mass function estimators, they have the advantage that the $\Phi$ values in different mass bins are not independent form each other. This makes these two techniques more robust towards deviations in the mass completeness function. For a more detailed discussion of each of the mass function estimators, their advantages and disadvantages and systematics that might affect them see \cite{Weigel:2016aa}. 

\section{Major merger and post merger stellar mass functions}
In the following section we present the stellar mass functions of local major mergers and post mergers. We use these stellar mass functions to determine the major merger fraction as a function of stellar mass. Furthermore, we determine the stellar mass functions of major mergers in different environments to compare our results to the predictions by \citetalias{Peng:2010aa}.
\subsection{Stellar mass functions of major mergers and post mergers}\label{sec:smf_mm_pm}
\begin{figure}
	\includegraphics[width=\columnwidth]{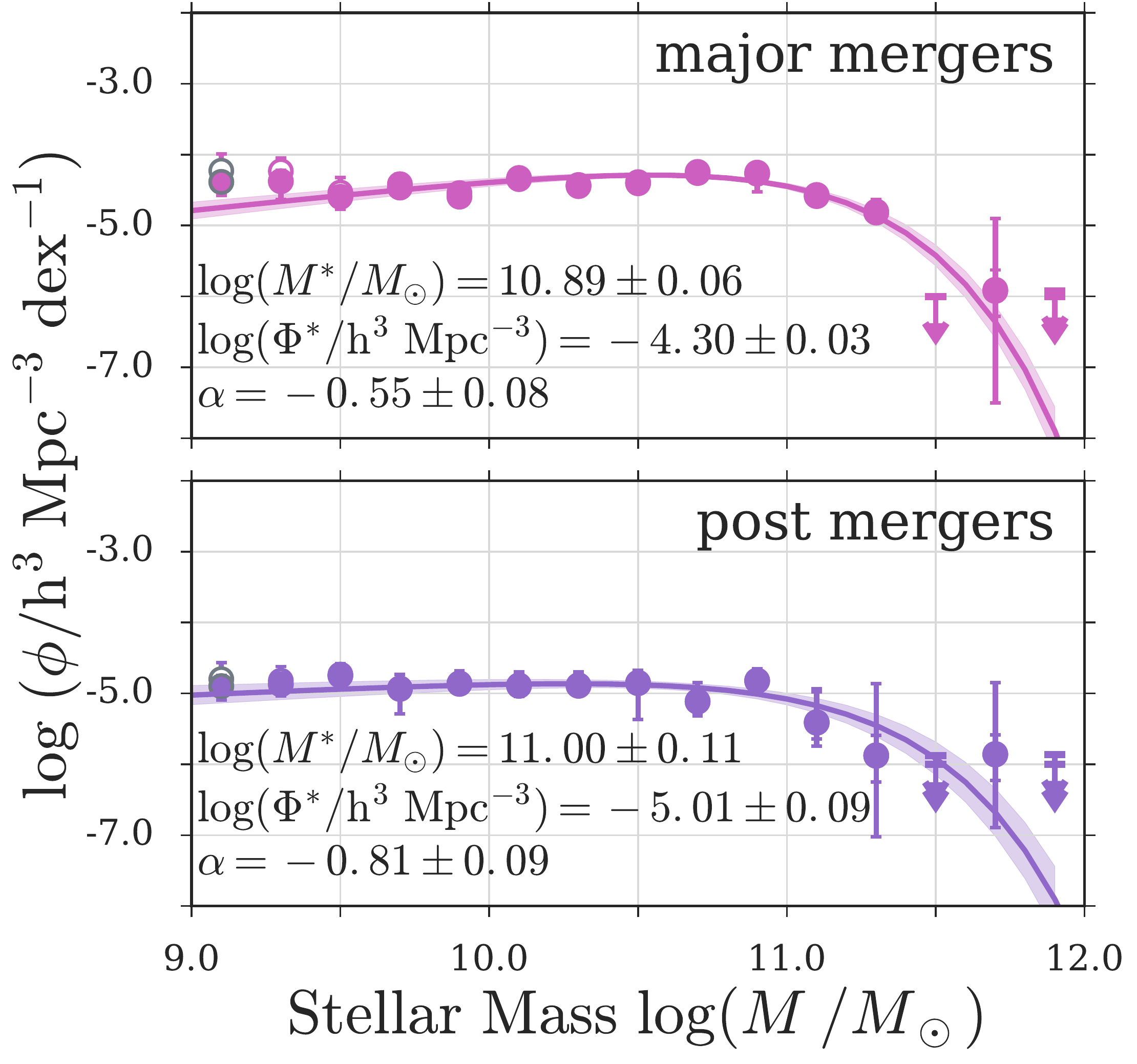}
	\caption{\label{fig:merger_post_merger_smf} Stellar mass functions of local major mergers and post mergers. To construct these stellar mass functions we use the samples of visually classified  major mergers and post mergers by  \protect\citetalias{Darg:2010aa} and \protect\cite{Carpineti:2012aa}, respectively. The open (1/\Vmax) and filled (SWML) symbols and solid lines (STY) show the results of different stellar mass function techniques according to \protect\cite{Weigel:2016aa}. Upper limits are computed based on both the 1/\Vmax\ and SWML technique. The best-fitting Schechter function parameters are given within the panels and are summarized in Table \ref{tab:parameters}.} 
\end{figure}

\begin{deluxetable*}{l|lllllllll}
	\tablecolumns{10}
	\tablecaption{\label{tab:parameters}Best-fitting Schechter function parameters}
	\tablehead{
		\colhead{sample} &
		\colhead{S} & 
		\colhead{D} & 
		\colhead{Nr. objects} & 
		\colhead{$\log (M^{*}/M_{\odot})$} & 
		\colhead{$\log (\Phi^{*}/\rm h^3 Mpc^{-3})$} & 
		\colhead{$\alpha$} & 
		\colhead{$\log (\Phi^{*}_2/\rm h^3 Mpc^{-3})$} & 
		\colhead{$\alpha_2$} & 
		\colhead{$\chi^{2}_{\rm reduced}$}
		}
		\startdata
			{entire sample}  & {} & {X} & {69289} & {$10.79 \pm 0.01$} & {$-3.31 \pm 0.20$} & {$-1.69 \pm 0.10$} & {$-2.01 \pm 0.28$} & {$-0.79 \pm 0.04$} & {$9.78$}\\
			{blue}  & {X} & {} & {32825} & {$10.60 \pm 0.01$} & {$-2.43 \pm 0.01$} & {$-1.21 \pm 0.01$} & {$$} & {$$} & {$4.70$}\\
			{green}  & {} & {X} & {13429} & {$10.65 \pm 0.02$} & {$-3.95 \pm 0.23$} & {$-1.84 \pm 0.15$} & {$-2.54 \pm 0.33$} & {$-0.44 \pm 0.07$} &  {$5.94$}\\
			{red}  & {} & {X} & {26143} & {$10.77 \pm 0.01$} & {$-6.73 \pm 0.79$} & {$-3.12 \pm 0.51$} & {$-2.21 \pm 1.12$} & {$-0.46 \pm 0.02$}  & {$6.67$}\\
			{major mergers}  & {X} & {} & {276} & {$10.89 \pm 0.06$} & {$-4.30 \pm 0.03$} & {$-0.55 \pm 0.08$} & {$$} & {$$} &  {$1.93$}\\
			{post mergers}  & {X} & {} & {104} & {$11.00 \pm 0.11$} & {$-5.01 \pm 0.09$} & {$-0.81 \pm 0.09$} & {$$} & {$$} &  {$1.04$}\\
			{Early types \& blue}  & {X} & {} & {219} & {$10.66 \pm 0.08$} & {$-4.47 \pm 0.06$} & {$-0.72 \pm 0.11$} & {$$} & {$$} & {$0.72$}\\
			{Early types \& green}  & {} & {X} & {735} & {$10.75 \pm 0.05$} & {$-7.14 \pm 1.12$} & {$-2.95 \pm 0.71$} & {$-3.82 \pm 1.57$} & {$-0.46 \pm 0.16$} & {$0.74$}\\
			{Early types \& red}  & {} & {X} & {8035} & {$10.74 \pm 0.01$} & {$-7.07 \pm 0.80$} & {$-3.09 \pm 0.57$} & {$-2.62 \pm 1.14$} & {$0.13 \pm 0.03$} & {$3.19$}\\
			{major mergers  \& $\log(\delta + 1) > 0.05$}  & {} & {X} & {216} & {$10.77 \pm 0.08$} & {$-7.76 \pm 1.12$} & {$-2.99 \pm 0.70$} & {$-4.25 \pm 1.58$} & {$-0.14 \pm 0.23$} &  {$1.36$}\\
			{major mergers \& $\log(\delta + 1) \leq 0.05$}  & {X} & {} & {68} & {$10.90 \pm 0.16$} & {$-5.25 \pm 0.15$} & {$-0.92 \pm 0.13$} & {} & {} &  {$1.06$}\\
			{mergers \& satellites}  & {X} & {} & {134} & {$10.94 \pm 0.10$} & {$-4.70 \pm 0.07$} & {$-0.69 \pm 0.11$} & {} & {} &  {$1.84$}\\
			{mergers \& centrals}  & {X} & {} & {161} & {$10.90 \pm 0.08$} & {$-4.56 \pm 0.04$} & {$-0.53 \pm 0.10$} & {} & {} & {$2.22$}\\
		\enddata
		\tablecomments{We determine the parameters based on the parametric maximum likelihood approach (STY, \protect\citealt{Sandage:1979aa}) and give the $1\sigma$ random errors which we compute directly from the STY MCMC chain. The second and third columns show if the subsample is better described by a single (S) or by a double (D) Schechter function according to the likelihood ratio test, which we use to compare the STY single and double Schechter likelihoods. The number of objects given in the fourth column corresponds to the number of galaxies above the mass completeness cut. The $\chi^2_{\rm reduced}$ value given in the last column was derived by comparing the non-parametric maximum likelihood values (SWML, \protect\citealt{Efstathiou:1988aa}) to the STY best-fitting Schechter function.}
\end{deluxetable*}

In Fig. \ref{fig:merger_post_merger_smf} we show the stellar mass functions of local major mergers and post mergers based on the samples by \citetalias{Darg:2010aa} and \cite{Carpineti:2012aa}. As we discussed above, we combine three stellar mass function techniques according to \cite{Weigel:2016aa}. The results of the classical 1/\Vmax\ and the SWML method are shown with open and filled symbols, respectively. The solid lines illustrate the results of the STY technique, the shaded regions show the corresponding $1 \sigma$ error contours. Upper limits are computed and shown for the 1/\Vmax\ and SWML results. The best-fitting Schechter function parameters are given within the panels and are also given in Table \ref{tab:parameters} which summarizes the parameters of all stellar mass functions used in this analysis.
\subsection{Merger fraction}\label{sec:merger_frac}
\begin{figure*}
	\includegraphics[width=\textwidth]{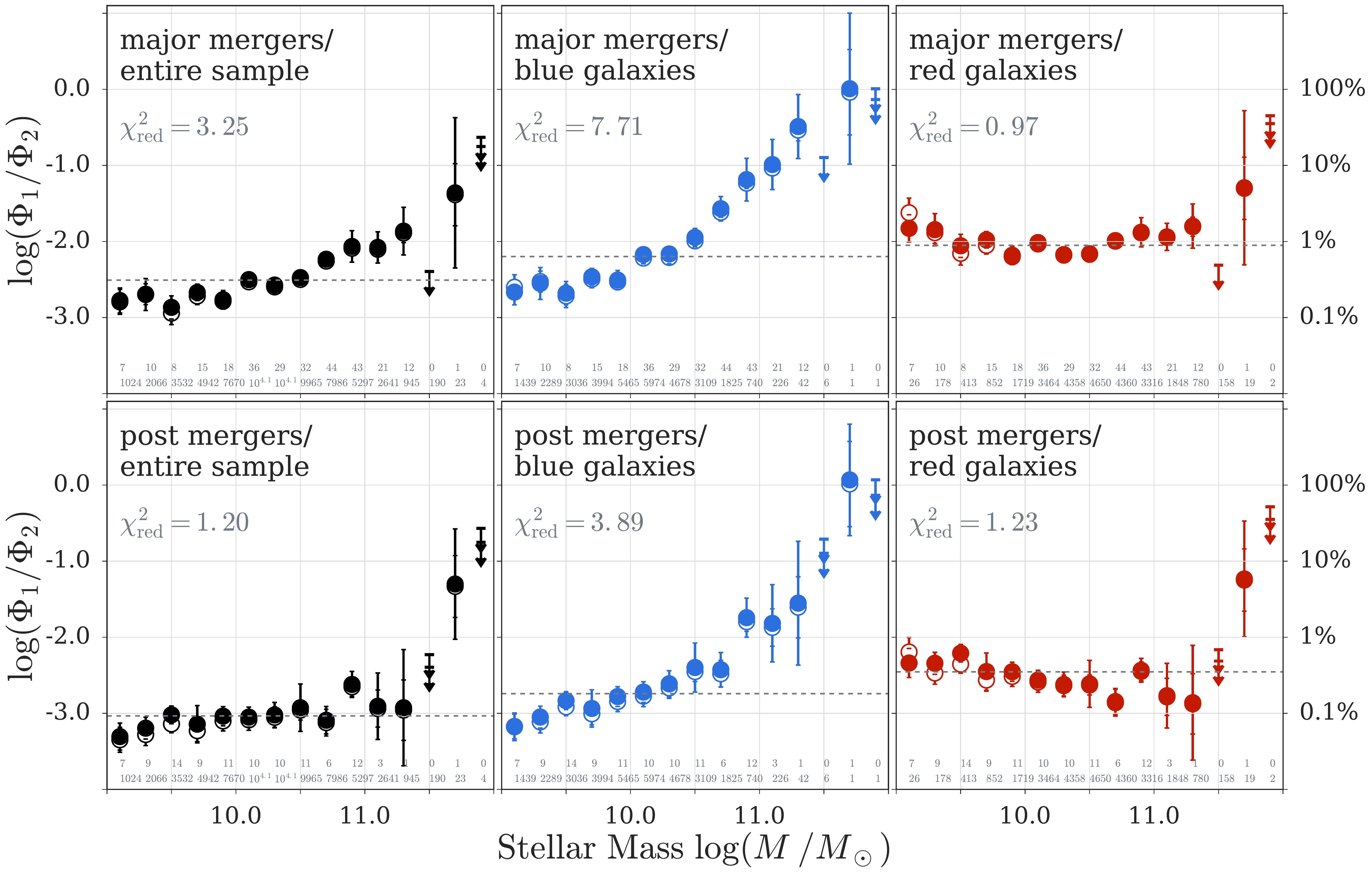}
	\caption{\label{fig:fraction}Merger fractions. The top panels show the number density of major mergers relative to the number density of all, all blue and all red galaxies. The bottom panels show the number of post mergers relative to the entire galaxy sample, all blue galaxies and all red galaxies. In analogy to Fig. \ref{fig:merger_post_merger_smf}, the open and filled markers show the fractions based on the 1/\Vmax\ and SWML results, respectively. Upper limits are computed using the $\Phi$ values from both methods which is why for some of the mass bins we show two upper limits. The grey dashed lines show the best-fitting relation for a constant fraction. The corresponding $\chi^2_{\rm reduced}$ values are given within the panels. The numbers at the bottom of the panels correspond to the number of galaxies in each mass bin that were used to compute the stellar mass functions. The top and the bottom row show the number of objects used for the numerator and the denominator mass functions, respectively.}
\end{figure*}

Having determined the stellar mass functions for major mergers and post mergers, we are able to constrain the merger and post merger fraction as a function of stellar mass. In the top panels of Fig. \ref{fig:fraction} we compare the number densities of major mergers to the entire galaxy sample, all blue galaxies and all red galaxies. The bottom panels illustrate the number density of post mergers relative to the entire galaxy sample, all blue galaxies and all red galaxies. At the bottom of each panel we give the number of objects in each mass bin which were used to compute the stellar mass functions. The top and the bottom row show the number of objects used for the numerator and the denominator, respectively. 

We compare the major merger fraction relative to all galaxies in our sample (top left-hand panel in Fig. \ref{fig:fraction}) to the results by \citetalias{Darg:2010aa}. \citetalias{Darg:2010aa} report a major merger fraction that ranges between $1.5 - 4.5\%$. This estimate is not based on the original major merger catalog, which we use for our analysis, but on an extended sample of `strongly perturbed' galaxies in the local Universe. \citetalias{Darg:2010aa} add expert visual assessments of galaxies with $f_{\rm m} < 0.4$ to derive the overall fraction of galaxies observed in a major merger. Furthermore, \citetalias{Darg:2010aa} introduce an absolute magnitude limit ($M_{\rm r} < -20.55$) to construct a volume complete sample. For our sample the number of major mergers relative to all galaxies ranges from $0 - 10\%$ at a given stellar mass. There is no need to construct a volume complete sample since we use stellar mass functions to measure the merger fraction, i.e. we correct for volume and stellar mass completeness effects. By integrating the merger fraction over $M$ from $9 < \log(M/M_\odot) < 12$ (not including upper limits) we find a fraction of $\sim 2\%$. Note that our stellar mass functions show the number density of major merger systems and not the number density of galaxies involved in a major merger. Assuming that on average each merger system contains two galaxies, the fraction of galaxies in a major merger relative to all galaxies is thus $\sim 4\%$ which is consistent with the results by \citetalias{Darg:2010aa}. 

Fig. \ref{fig:fraction} shows an increase in the major merger fraction relative to all blue galaxies towards higher masses. A similar, but weaker trend can be seen for the number of major mergers relative to the entire galaxy sample and relative to red galaxies. 

For simulations this trend has been discussed by, for example,  \cite{Bertone:2009aa} and \cite{Hopkins:2010aa,Hopkins:2010ab}. \cite{Hopkins:2010aa,Hopkins:2010ab} argue that while the halo merger fraction shows no strong halo mass dependence, it is the stellar mass to halo mass conversion (e.g. \citealt{Behroozi:2013aa}) that introduces the stellar mass dependence in the galaxy merger fraction. At low halo masses, a 1:3 halo mass merger corresponds to a minor galaxy merger since the stellar mass to halo mass relation is steep, i.e. a small halo mass range corresponds to a wide stellar mass range. At high halo masses, even a minor halo mass merger corresponds to a major galaxy merger as the stellar mass to halo mass conversion is shallow, i.e. a wide range in halo mass corresponds to a small range in stellar mass. Compared to the halo major merger fraction, the galaxy major merger fraction is thus suppressed at low stellar masses and enhanced at high stellar masses.

Observational estimates of the merger fraction are method dependent \citep{Lotz:2011aa}. Due to this, no clear consensus regarding the mass or luminosity dependence of the major merger fraction has been reached. For example,  \cite{Casteels:2014aa} use morphological measurements based on concentration, asymmetry and clumpiness (CAS) to identify major mergers. They find a merger fraction that is consistent with being constant at stellar masses $9.5 < \log (M/M_\odot) < 11.5$. This is in agreement with the results of \cite{Xu:2012aa} who find a constant pair fraction for the same mass range. At stellar masses below $\log(M/M_\odot) = 9.5$ \cite{Casteels:2014aa} find an increased merger fraction.  \cite{Domingue:2009aa} and \cite{Xu:2004aa} find close pair fractions that are constant and increase with luminosity, respectively. At $z\sim0.5$ \cite{Bundy:2009aa} find a pair fraction that increases as a function of stellar mass. 

\subsection{Mass and environment quenching}
\label{sec:mass_env_qu}
\begin{figure*}
	\includegraphics[width=\textwidth]{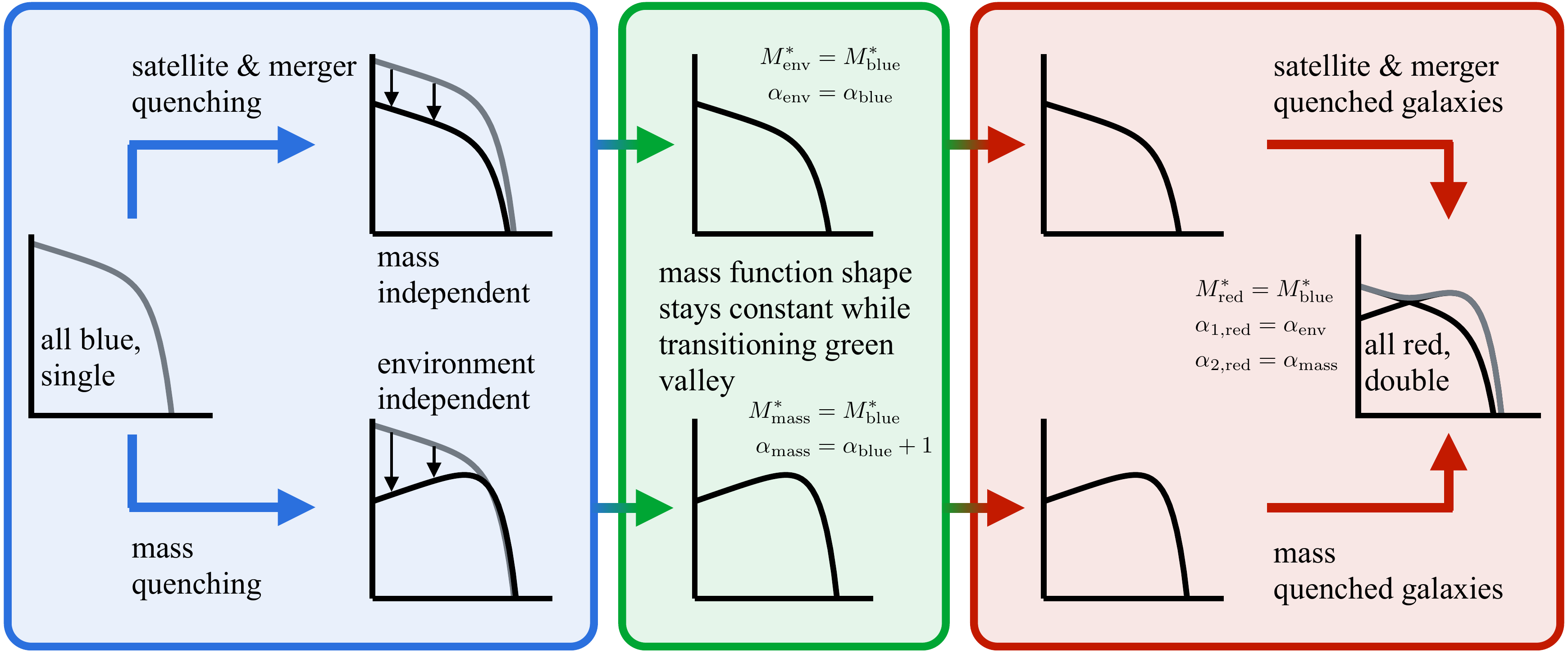}
	\caption{
		\label{fig:peng_cartoon} Schematic figure illustrating the phenomenological model by \protect\citetalias{Peng:2010aa}. In their model \protect\citetalias{Peng:2010aa} use three different quenching mechanisms to explain the double Schechter shape of the stellar mass function of red galaxies. \textit{Mass quenching} (bottom row) is a mass dependent, but environment independent process. The cause of mass quenching is most likely an internal process, such as AGN feedback (e.g. \protect\citealt{Fabian:2012aa}) or secular processes (e.g. \protect\citealt{Masters:2011aa}). The probability of a galaxy being mass quenched increases as a function of its stellar mass. \protect\citetalias{Peng:2010aa} thus propose that when selecting blue galaxies that are in the process of being mass quenched, we will observe a stellar mass function that has the same $M^{*}$ as blue galaxies, but a shallower slope $\alpha$ ($M^{*}_{\rm mass} = M^{*}_{\rm blue}$, $\alpha_{\rm mass} = \alpha_{\rm blue} + 1$). In the \protect\citetalias{Peng:2010aa} model \textit{satellite} and \textit {merger quenching} (top row) are processes that are mass independent, but environment dependent. According to \protect\citetalias{Peng:2010aa} these processes are different manifestations of a dark matter halo merger. Galaxies that are being satellite or merger quenched have the shape of the blue, star forming mass function ($M^{*}_{\rm env} = M^{*}_{\rm blue}$, $\alpha_{\rm env} = \alpha_{\rm blue}$). Leading to the same mass function shapes, these effects can thus be summarized as one environmental quenching process. While transition the green valley galaxies do not gain significant amounts of mass. The stellar mass functions hence retain their shapes. Mass, merger and satellite quenched galaxies make up the red sequence. The double Schechter shaped red mass function is the combination of the mass and environment quenched single Schechter mass functions ($M^{*}_{\rm red} = M^{*}_{\rm blue}$, $\alpha_{1, \rm red} = \alpha_{\rm env}$, $\alpha_{2, \rm red} = \alpha_{\rm mass}$). 
	}
\end{figure*}

\begin{figure*}
	\includegraphics[width=\textwidth]{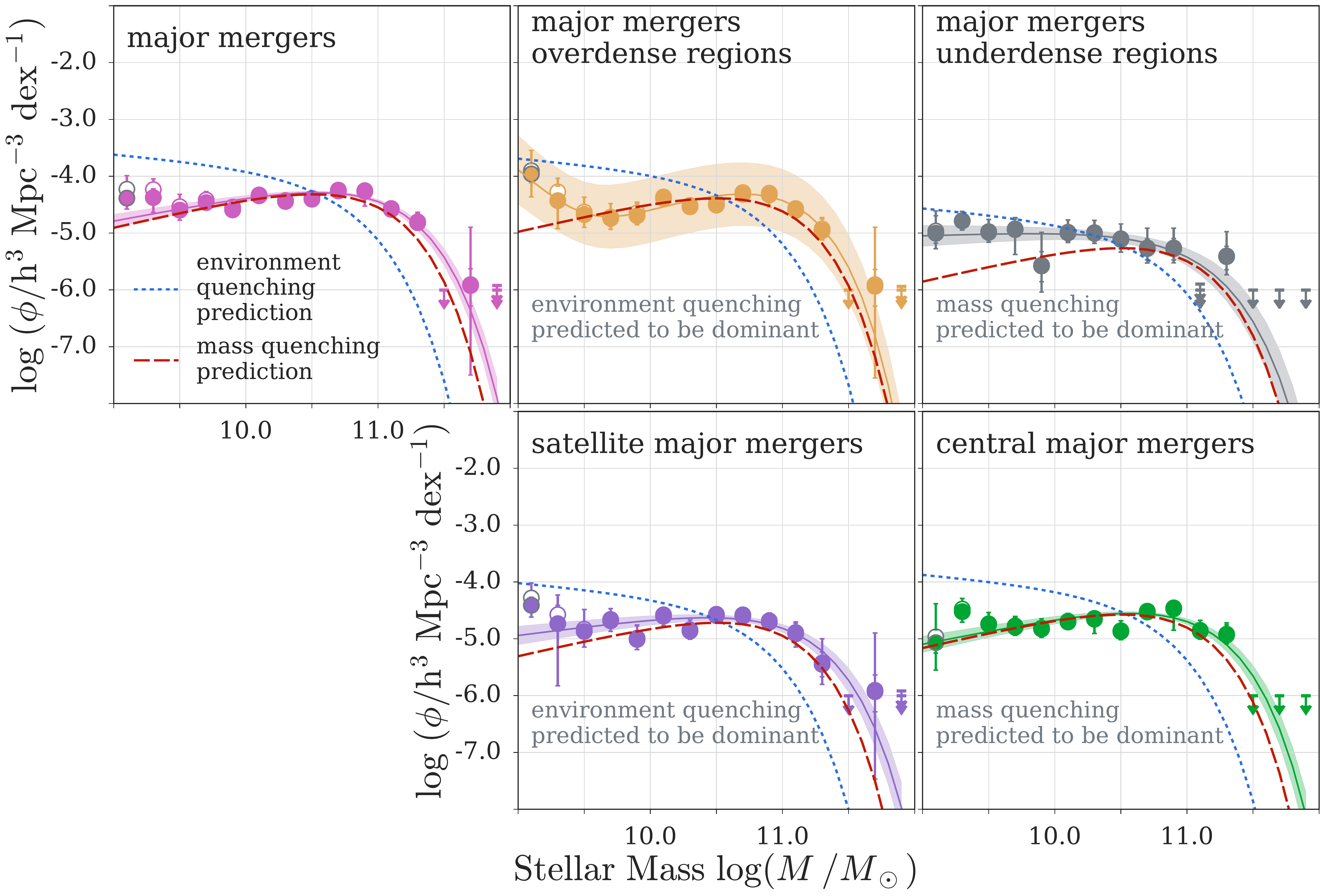}
	\caption{\label{fig:env_dens}Stellar mass functions for major mergers, major mergers in overdense and underdense regions and satellite and central major mergers. An important part of the empirical  \protect\citetalias{Peng:2010aa} model is the fact that mass and environment dependent quenching mechanisms can be disentangled by considering over- and underdense regions. Environment quenching is dominant in overdense regions and mostly affects satellites. Mass quenching becomes apparent in underdense regions and primarily quenches centrals. We split the major merger sample by density \protect\citep{Weigel:2016aa} and into centrals and satellites \protect\citep{Yang:2007aa} to test if the major merger mass function is consistent with the \protect\citetalias{Peng:2010aa} model. From top left to bottom right, we show the stellar mass functions of all major mergers, mergers in overdense and underdense regions and satellite and central major mergers. Overplotted in red and blue are the predictions for mass and environment quenching according to the \protect\citetalias{Peng:2010aa} model. To allow for an easier comparison the predicted mass functions are rescaled to have the same $\Phi^{*}$ as the observed ones. In the \protect\citetalias{Peng:2010aa} model merger quenching is an environment dependent, but mass independent process. The observed mass function of all major mergers (top left-hand plot) should thus resemble the predicted environment quenching mass function (blue dotted line). Yet contrary to the  \protect\citetalias{Peng:2010aa} prediction, the merger mass function shape is consistent with the mass quenching mass function. The shapes of the observed merger mass functions in over- and underdense regions neither match the predicted mass nor environment quenching mass functions.  As predicted by \protect\citetalias{Peng:2010aa}, the observed mass function of central major mergers is consistent with the predicted mass quenching mass function. For satellite major mergers environmental quenching effects should be dominant. However the slope $\alpha$ of the observed satellite major merger mass function is too shallow to be consistent with environment quenching. This figure thus illustrates that our observations of the merger quenching process in the local Universe are inconsistent with the empirical model by \protect\citetalias{Peng:2010aa}.}
\end{figure*}

In the following section we compare the major merger stellar mass function to the predictions by \citetalias{Peng:2010aa}. We first summarize the results by \citetalias{Peng:2010aa} and then discuss the implications of our measurements.

\subsubsection{The empirical model}
In their purely empirical model \citetalias{Peng:2010aa} consider three physical processes that are likely to lead to quenching and predict the corresponding stellar mass functions. The following three processes are the quenching channels that they consider:

\begin{itemize}
	\item \textit{mass quenching}: Mass quenching is independent of the environment, but does depend on stellar mass. Mass quenched galaxies follow a single Schechter function. Compared to the mass function of blue galaxies, this Schechter function has the same $M^{*}$, but a shallower, more positive slope $\alpha$ ($M^{*}_\mathrm{mass} = M^{*}_\mathrm{blue}$, $\alpha_\mathrm{mass} = \alpha_{\rm blue} + 1$). Mass quenching could be associated with AGN feedback \citep{Silk:1998aa, Di-Matteo:2005aa,Schawinski:2006aa, Kauffmann:2007aa, Georgakakis:2008aa, Hickox:2009aa, Cattaneo:2009ab, Fabian:2012aa, Bongiorno:2016aa, Smethurst:2016aa} or secular processes \citep{Kormendy:2004aa, Masters:2011aa, Cheung:2013aa}.
	
	\item \textit{satellite quenching}: Satellite quenching is mass independent, but environment dependent. As the satellite quenching efficiency is mass independent, the stellar mass function of satellite quenched galaxies has the same single Schechter function shape as blue, star forming galaxies ($M^{*}_\mathrm{env} = M^{*}_\mathrm{blue}$, $\alpha_\mathrm{env} = \alpha_\mathrm{blue}$). Satellite quenching could be associated with external processes such as ram pressure stripping \citep{Gunn:1972aa} or strangulation \citep{Larson:1980aa, Balogh:2000aa}.
	
	\item \textit{merger quenching}: Merger quenching has the same properties as satellite quenching. \citetalias{Peng:2010aa} assume that the merger quenching efficiency is mass independent, but environment dependent. They thus predict the mass function of merger quenched galaxies to have the same shape as the mass function of satellite quenched galaxies. 
\end{itemize}

The sum of mass, satellite and merger quenched galaxies makes up the red sequence and the combination of their respective single Schechter functions reproduces the double Schechter function that we observe for red galaxies ($M^{*}_{\rm red} = M^{*}_{\rm blue}$, $\alpha_{1, \rm red} = \alpha_{\rm env}$, $\alpha_{2, \rm red} = \alpha_{\rm mass}$). We illustrate the effect of mass, satellite and merger quenching in Fig. \ref{fig:peng_cartoon}.

As merger and satellite quenching have the same properties and result in the same stellar mass function shape, these two processes are often considered as one environment dependent quenching channel. \citetalias{Peng:2010aa} argue that merger and satellite quenching are different manifestations of the same physical processes: a dark matter halo merger. If the baryonic galaxies merge we observe a merger quenched galaxy. If the baryonic galaxies do not merge we conclude that satellite quenching has occurred. We refer to these two processes as environment quenching. 

An important part of the \citetalias{Peng:2010aa} model is the fact that the quenching channels can be disentangled based on their environmental dependence. According to \citetalias{Peng:2010aa}, satellite and merger quenching is dominant in overdense regions and mostly affects satellites. Mass quenching becomes apparent in underdense regions and mainly causes centrals to quench.

\subsubsection{The observations}

We now compare our major merger mass function to the predictions by \citetalias{Peng:2010aa}. We not only consider the major merger mass function, but also split the major merger sample by environmental density to compare to the \citetalias{Peng:2010aa} predictions for different environments.

We determine the mass function of major mergers in over- and in underdense regions \citep{Weigel:2016aa} and use the central/satellite classification by \cite{Yang:2007aa} to split the major merger sample into centrals and satellites. Note that this does not imply that we are generating the mass functions of major satellite - satellite or major central - central mergers. Instead, we determine if the galaxies which we are considering in the construction of the major merger mass function are classified as a satellite or as a central (see Section \ref{sec:merger_sample}).  

We show the stellar mass functions for major mergers, major mergers in over and underdense regions and the mass functions of central and satellite major mergers in Fig. \ref{fig:env_dens}. Overplotted with dashed and dotted lines, we illustrate the shape that we would expect to see for mass and environment quenching, respectively. Note that \citetalias{Peng:2010aa} only predict the shape and not the normalization of the mass and environment quenching mass functions. For an easier comparison we thus rescale the predicted mass functions to match the $\Phi^{*}$ of the subsample that we are considering. 

First, let us consider the general major merger mass function which is shown in the upper left-hand panel of Fig. \ref{fig:env_dens}. According to the \citetalias{Peng:2010aa} model merger quenching is an environment dependent effect. The mass function of merger quenched galaxies should thus have the same shape as the blue mass function, indicated by the blue dotted line in Fig. \ref{fig:env_dens}. Contrary to the prediction by \citetalias{Peng:2010aa}, the observed major merger mass function is however consistent with the mass quenching mass function.  

We find a similar inconsistency if we split the major merger sample by overdensity. The observed stellar mass functions for mergers in over- and underdense regions are shown in the upper middle and right-hand panels, respectively. The mass function of mergers in overdense regions neither resembles the mass nor the environment quenching mass function. In underdense regions, mass quenching effects should be dominant. Yet the observed merger mass function has an $M^{*}$ that is too high to be consistent with the environment quenching mass function and is too flat to follow the mass quenching mass function.  

The mass function of central major mergers, shown in the bottom right-hand panel, is the only case where our observations match the expectations. As predicted, the mass function of central major mergers is consistent with the mass quenching mass function. The mass function of satellite major mergers does however also resemble the mass quenching mass function, even though environmental quenching effects should dominate. This is shown in the middle panel at the bottom. 

We conclude that our measurement of the major merger mass function is inconsistent with the empirical model by \citetalias{Peng:2010aa}. Besides the major merger mass functions, there are additional indicators of the \citetalias{Peng:2010aa} model being over simplified with respect to merger quenching. A fundamental assumption of the model is the fact that the mass functions of red and blue galaxies have the same $M^{*}$. However for our sample $M^{*}_{\rm red} - M^{*}_{\rm blue} \sim 0.2\ \rm dex$ (see Table \ref{tab:parameters}, \citealt{Weigel:2016aa}). Furthermore, \citetalias{Peng:2010aa} assume a mass independent merger quenching efficiency. Contrary to this assumption, we find a mass dependent merger fraction, as we discussed in Section \ref{sec:merger_frac} and show in Fig. \ref{fig:fraction}. \citetalias{Peng:2010aa} also assume that the merger quenching efficiency increases as a function of environmental density. Yet in massive systems such as clusters, galaxies have high relative velocities and the probability of mergers is expected to decrease \citep{Ostriker:1980aa}.

\section{Major mergers as a quenching mechanism - model}\label{sec:method}
\subsection{Assumptions and expectations}\label{sec:assumptions}
We now use the local major merger and post merger mass functions to investigate the process of major merger quenching. To do so we make the following straight forward assumptions:
\begin{enumerate}
	\item galaxies that are in the process of being major merger quenched evolve along the following sequence of stages which we refer to as the `merger quenching sequence': major merger, post merger, blue early type, green early type, red early type; 
	\item the probability of galaxies evolving from the major merger to the red early type stage is mass independent;
	\item while transitioning, the population of merger quenched galaxies does not increase its stellar mass significantly, thereby retaining its mass distribution. 
\end{enumerate}
These assumptions imply that the stellar mass functions of major mergers, post mergers, blue early types, green early types and red early types are similar in shape. They allow us to investigate: 
\begin{enumerate}
	\item {if major mergers are likely to lead to quenching,}
	\item{the relative amount of time spent in stages along the merger quenching sequence,}
	\item{and the significance of major merger quenching.}
\end{enumerate}

In the following section we motivate our assumptions. We discuss the merger quenching sequence and the order of its stages, the mass dependence of the major merger to red early type transition probability and the possible increase in stellar mass along the sequence. In Section \ref{sec:analysis} we apply these assumptions to our sample and use them to investigate major merger quenching.

\subsection{Major merger quenching stages and their order}\label{sec:theory}
We focus on mergers between gas-rich galaxies of comparable mass (mass ratio 1:3 and greater). According to \cite{Toomre:1972aa}  these major mergers are capable of transforming disc galaxies into spheroids or ellipticals which has now also been shown and studied in various simulations (e.g. \citealt{Mihos:1996aa, Springel:2005ad, Di-Matteo:2005aa, Hopkins:2006aa, Croton:2006aa, Hopkins:2008aa,Hopkins:2008ab, Khalatyan:2008aa, Somerville:2008aa}). Furthermore, previous studies have investigated the evolution of merging galaxies both in terms of colour (e.g. \citealt{Kaviraj:2011aa})  and SFR (e.g. \citealt{Springel:2005ad, Hopkins:2008ab}). Based on these studies we assume that galaxies which evolve along the classical \cite{Sanders:1988aa} quenching sequence are likely to pass through the following stages: 

\begin{enumerate}
	\item {\textit{major merger stage:}
		We base our analysis on the Galaxy Zoo major merger sample (\citetalias{Darg:2010aa}, \citealt{Darg:2010ab}) which contains visually classified merger systems. These systems consist of at least two strongly perturbed, close-by  galaxies. Disrupted tidal fields and dynamical friction drive the merging galaxies towards each other, violent relaxation re-arranges the stellar orbits \citep{Bournaud:2011ab}.    
	}
	\item {\textit{post merger stage:}
		After coalescence, only one nucleus remains and the galaxy is likely to have a disturbed morphology, it might for example be exhibiting tidal tails. Gravitational torques cause angular momentum loss and allow the gas to fall towards the center of the newly formed galaxy \citep{Mihos:1996aa, Hernquist:1989aa, Carpineti:2012aa}. The high central gas densities trigger a starburst and an AGN. Due to the large amounts of gas and dust the galaxy is classified as an ultraluminous infrared galaxy (ULIRG; \citealt{Sanders:1996aa, Genzel:2001aa}). 
	}
	\item {\textit{blue early type stage:}
		The galaxy has now lost its signs of a recent major merger (see discussion below) and appears to have an early type morphology. Kinetic and thermal feedback from the AGN and/or from supernovae expels or heats the gas in the galaxy, thereby quenching star formation \citep{Kaviraj:2007ab}. This has been predicted theoretically \citep{Di-Matteo:2005aa, Springel:2005ac, Croton:2006aa, Khalatyan:2008aa, Somerville:2008aa} and confirmed observationally \citep{Schawinski:2006aa, Tremonti:2007aa, Schawinski:2007aa, Wong:2015aa}. 
	}
	\item {\textit{green early type stage:}
		As the SFR declines, the galaxy transitions through the green valley. Showing signs of recent star formation, the galaxy is classified as a  post-starburst galaxy (PSG/E+A/K+A; \citealt{Bekki:2001aa, Goto:2005aa, Yamauchi:2008aa, Wong:2012aa}). This stage can also be accompanied by AGN activity \citep{Yan:2006aa}.  				
	}
	\item {\textit{red early type stage:}
		Once the remaining gas is consumed, the galaxy reddens and reaches the red sequence as a red early type galaxy.}
\end{enumerate}

The sequence that we set out above consists of dividing the entire galaxy sample both in terms of morphology and colour. First, we use visual morphologies to select major mergers, post mergers and early types. Second, we use the optical colour as a proxy for SFR to select early types in the blue cloud, green valley and red sequence. This allows us to trace the shut down of star formation in these merger remnants.

Along the sequence galaxies transition from the post-merger to the blue early type stage. We thus seem to assume that the change in morphology precedes the change in colour. This seems to imply that the dynamical or the relaxation time scale of merger remnants is shorter than the duration of their starbursts. Yet it is important to consider the data that we will be applying this model to. Specifically, the selection of early type galaxies has to be taken into account. 

In Section \ref{sec:analysis} we will apply our assumptions to a sample of galaxies that have been classified by Galaxy Zoo users. Users were asked to classify galaxies according to their SDSS images.  \cite{Schawinski:2010aa} use a  sample of blue early type galaxies with SDSS classifications \citep{Schawinski:2007aa} to show that at least $50\%$ of all blue early types show signs of a recent merger in co-added Stripe 82 images. These images are approximately two magnitudes deeper than regular SDSS images. Similarly,  \cite{van-Dokkum:2005ab} uses deep imaging ($\sim 28\ \rm mag\ arcsec^{-2}$) from the Multi-wavelength Survey by Yale-Chile (MUSYC; \citealt{Gawiser:2006aa}) and  the NOAO Deep Wide-Field Survey (NDWFS; \citealt{Jannuzi:1999aa}) and finds that $53\%$ of the nearby, red galaxies in the sample show signs of tidal interactions. When restricting the sample to bulge-dominated early type galaxies this fraction increases to $71\%$. We are thus not proposing that merger remnants have lost all signs of recent mergers and are fully relaxed by the time they leave the blue cloud. Instead we assume that the low surface brightness tidal features have faded and are no longer visible in the shallow SDSS images. Missing the signs of morphological disturbance, the still blue merger remnants are classified as early types.   

We also note that AGN activity has been found to peak during different stages along the merger quenching sequence \citep{Schawinski:2010aa, Koss:2010aa, Ellison:2011aa, Carpineti:2012aa, Kaviraj:2015aa, Carpineti:2015aa}. As we discussed above the definition of the post merger and blue early type stage depends on the specifics of the sample and the method used to select, for instance, post merger galaxies. It is thus difficult to directly compare previous studies. In the sequence that we set out above, black holes increase their accretion rate during the post merger stage and provide the necessary feedback for a decrease in SFR in the blue early type stage. The result of AGN and star formation feedback becomes apparent during the green early type stage, once the galaxy has significantly decreased its SFR. However, we cannot determine the exact time during the evolution at which galaxies have experienced sufficient feedback to quench their star formation.

\subsection{Mass dependence of the transition probability}\label{sec:mass_dep_prob}
Below we discuss three processes that could cause the probability of a galaxy to transition from a major merger to a red early type stage to be mass dependent. As stated above, we assume that these effects are negligible and do not introduce a significant mass dependence in the space density of galaxies that are being major merger quenched.

\begin{enumerate}
	\item {\textit{reforming of a disc:}
		Simulations have shown that, unlike the sequence that we laid out above, a major merger between two spiral galaxies of comparable mass can also lead to the formation of a new spiral galaxy \citep{Hernquist:1991aa, Barnes:1996aa, Barnes:2002aa, Naab:2006aa, Robertson:2008aa}. In most models the probability of reforming a disc depends on the gas fraction within the merging galaxies and the amount of stellar or AGN feedback during the merger. Other parameters such as the mass ratio of the merging galaxies, their orbital parameters and their mass distributions also affect the probability of regrowing a disc (see e.g. \citealt{Hopkins:2009ac}). The high gas fractions that are necessary for a merger remnant to be able to regrow its disc are typically found in galaxies at $z>1$ \citep{Daddi:2010aa, Tacconi:2010aa}. In the model by \cite{Robertson:2006aa} the gas fraction has to be above $50\%$ for a disc to reform. \cite{Springel:2005ab} use pure gas discs to model the reformation of a rotationally supported disc after the merger. In the model by \cite{Governato:2009aa} the merging galaxies have a gas fraction below $25\%$ at $z<3$, yet the simulation involves constant accretion of gas through cold streams and efficient gas cooling. 
		
		Observationally the regrowing of a disc in a post merger galaxy has been observed in the local Universe (see e.g. \citealt{Hau:2008aa, Kannappan:2009aa, Salim:2012aa, Moffett:2012aa, Ueda:2014ab, George:2017aa}). Blue early type galaxies with signs of a disc seem to be primarily occur in low mass galaxies (e.g. $< 3 \times 10^{10} M_{\odot}$ \citealt{Kannappan:2009aa}). Compared to more massive galaxies, the regrowing of a disc after a major merger event in lower mass galaxies could be promoted by higher gas fractions \citep{Catinella:2010aa}.
		
		The regrowing of a disc after a major merger event is hence theoretically possible and has been observed for a small sample of local galaxies. Yet the significance of this effect on the mass dependence of the merger quenching probability remains unclear. Robust to a mild mass dependence, the mass functions of galaxies along the merger quenching sequence would however only be affected if this is a strongly mass dependent effect. For instance, if $10^{9} M_{\odot}$ galaxies are more than $50\%$ more likely to regrow a disc than $M^{*}$ galaxies, the resulting difference in the space densities at these masses would affect the $\alpha$ of the resulting mass function. We make the simplified assumption that such an extreme effect is unlikely and neglect the process of disc reforming on the transition probability.}
	
	\item{\textit{AGN feedback:} 
		AGN feedback is necessary to efficiently quench a merger remnant \citep{Springel:2005ad, Birnboim:2007aa, Khalatyan:2008aa, Hopkins:2008ab}. Without AGN feedback a merger remnant can return to being a star forming late type \citep{Sparre:2016aa}. Furthermore, AGN feedback has to be introduced to explain the high gas depletion rate and the rapid early type evolution \citep{Schawinski:2014aa, Smethurst:2016aa} from blue to red \citep{Kaviraj:2011aa}. Recent work has shown that observed AGN luminosity functions are consistent with a mass independent AGN fraction and accretion rate distribution (\citealt{Aird:2012aa}, \ERDFpaper). We thus assume that the probability of a merger remnant being affected by sufficient AGN feedback to transition to the red sequence is mass independent.}
	
	\item{\textit{dynamical friction}: 
		\cite{Chandrasekhar:1943aa} introduced the concept of dynamical friction being mass dependent. More recently, \cite{Jiang:2008aa} (also see \cite{Jiang:2010aa}) have shown that the resulting mass dependence of the merger time scale is best expressed as $T \propto M_{\rm primary}/( M_{\rm secondary} \times \ln(1 + M_{\rm primary}/M_{\rm secondary}))$. Here $M_{\rm primary}$ and $M_{\rm secondary}$ refer to the more massive and less massive merging partner, respectively. Given our major merger definition this causes up to a factor of 1.5 difference in the merging time scale of merging systems depending on their mass ratio. \cite{Jiang:2008aa} consider the time between the secondary first crossing the dark matter virial radius of the primary and the the beginning of the coalescence as their merging time. We assume that the galaxies in the \citetalias{Darg:2010aa} sample, which show clear signs of interaction, are close to coalescence and neglect the mass ratio and thus mass dependence of dynamical friction.}
	
	\item{\textit{fading of merger features:}
		As we discussed above, we assume that the post merger stage is followed by the blue early type stage, as for most galaxies the signs of a recent merger will have faded enough to no longer be detected in the shallow SDSS images. We assume that there is no mass dependence in the time scale over which these tidal features fade. There might be galaxies for which the signs of a recent merger fade less quickly and which might be classified as post mergers instead of, for instance, green early types. However, if mass independent, this effect does not introduce a bias in the space density of galaxies which we observe along the merger quenching sequence.		
	}
\end{enumerate}

\subsection{Mass increase along the quenching sequence}\label{sec:mm_quenching}
During their evolution from blue to red merger quenched galaxies can gain stellar mass through three different channels:
\begin{enumerate}
	\item{\textit{through star formation:} while transitioning from the blue cloud to the red sequence, a galaxy retains low levels of SFR. However even if a galaxy would keep its pre-quenching SFR, the amount of gained stellar mass would be negligible compared to the galaxy's already existing stellar mass. For example, a $M^{*}$ galaxy of $10^{10.8} M_{\odot}$ \citep{Weigel:2016aa} has a SFR of $\sim 5 M_{\odot}/\rm yr$ \citep{Lilly:2013aa} if it is on the main sequence. For a constant SFR, this galaxy will increase its stellar mass by a factor of $1.4$ within $5\ \rm Gyr$. So, even if the SFR were to stay constant during the transition from blue to red, the galaxy would only increase its stellar mass by 0.1 dex if the transition takes $\sim 5\ \rm Gyr$.}
	\item{\textit{through a starburst during the ULIRG phase:} as we mentioned above, a major merger quenched galaxy is likely to experience a starburst while transitioning from blue to red. Yet similar to the argument in the first point, the galaxy will not gain significant amounts of stellar mass during the starburst phase (see e.g.  \citealt{Genzel:1998aa, Carpineti:2015aa}). \cite{Di-Matteo:2008aa}, for instance, find that strong starbursts are rare in the local Universe and that a merger triggered starburst results in a SFR that is enhanced by less than a factor of five. They also find a typical starburst duration of the order of $10^8\ \rm yr$. For a $M^{*}$ galaxy on the main sequence this implies a $< 0.02$ dex increase in stellar mass.}
	\item{\textit{through the mass of the merging partner:} through merging, a galaxy can increase its stellar mass by 0.3 dex at most. Assume we consider the blue spiral galaxy $M_1$ in our major merger mass function construction. The galaxy $M_1$ is merging with, $M_2$, must have $M_2 \leq M_1$, otherwise we would have considered $M_2$ when determining the stellar mass function. $M_1$'s mass increases by 0.3 dex if $M_1 = M_2$.  Mergers with low mass ratios are more common than mergers between galaxies of comparable mass \citep{Kaviraj:2014aa}}. The number of galaxies that double their stellar mass, i.e. increase their mass by 0.3 dex, is thus likely to be low.
\end{enumerate}

While transitioning from blue to red, galaxies are thus unlikely to gain significant amounts of mass. We expect galaxies in evolutionary stages between the blue cloud and the red sequence to have mass functions of the same shape if the transition probability is mass independent. The stellar mass function normalizations are expected to be the same if all galaxies spend the same amount of time in each stage and if all galaxies transition from one stage to the next. The mass function of the blue, still star forming galaxies has a different normalization since galaxies spend a certain amount of time in this stage before the cessation of their star formation. The same is true for the mass function of the red and dead galaxies since this is the end stage and galaxies will accumulate here. The shape of the red mass function is the same as that of the transitioning objects, if the quenching process we are considering is the only way to build up a red galaxy and there are no other quenching channels. If galaxies spend less time in a certain stage or only a fraction of galaxies has transitioned from the previous stage, $\Phi^{*}$ of this stage decreases. The shape of the mass function of one of the phases changes if the transition from the previous stage is mass dependent and for example more efficient at higher stellar masses. 

\section{Major mergers as a quenching mechanism - analysis}\label{sec:analysis}
Based on the assumptions that we introduced in the previous section, we now investigate the effect of quenching through major mergers in the local Universe. We introduce the stellar mass functions of galaxies along the merger quenching sequence and test if they are similar in shape. We then use these stellar mass functions to estimate the relative amount of time that galaxies spent in stages along the sequence. To determine the significance of major merger quenching we introduce four tests which vary in their level of sophistication and assumptions. First, we compare the shapes of the major merger and the red early type mass functions. Second, we compare the major merger to the mass function of all green galaxies. Third, we estimate the contribution of major merger quenched galaxies to the green valley flux. Fourth, we simulate the evolution of the red stellar mass function and determine the fraction of galaxies that are likely to have been major merger quenched within the last 5 Gyr. We end this section by summarizing our results regarding the significance of major merger quenching. 
\subsection{Major merger quenching sequence mass functions}
\label{sec:mm_mf}
\begin{figure*}
	\includegraphics[width=\textwidth]{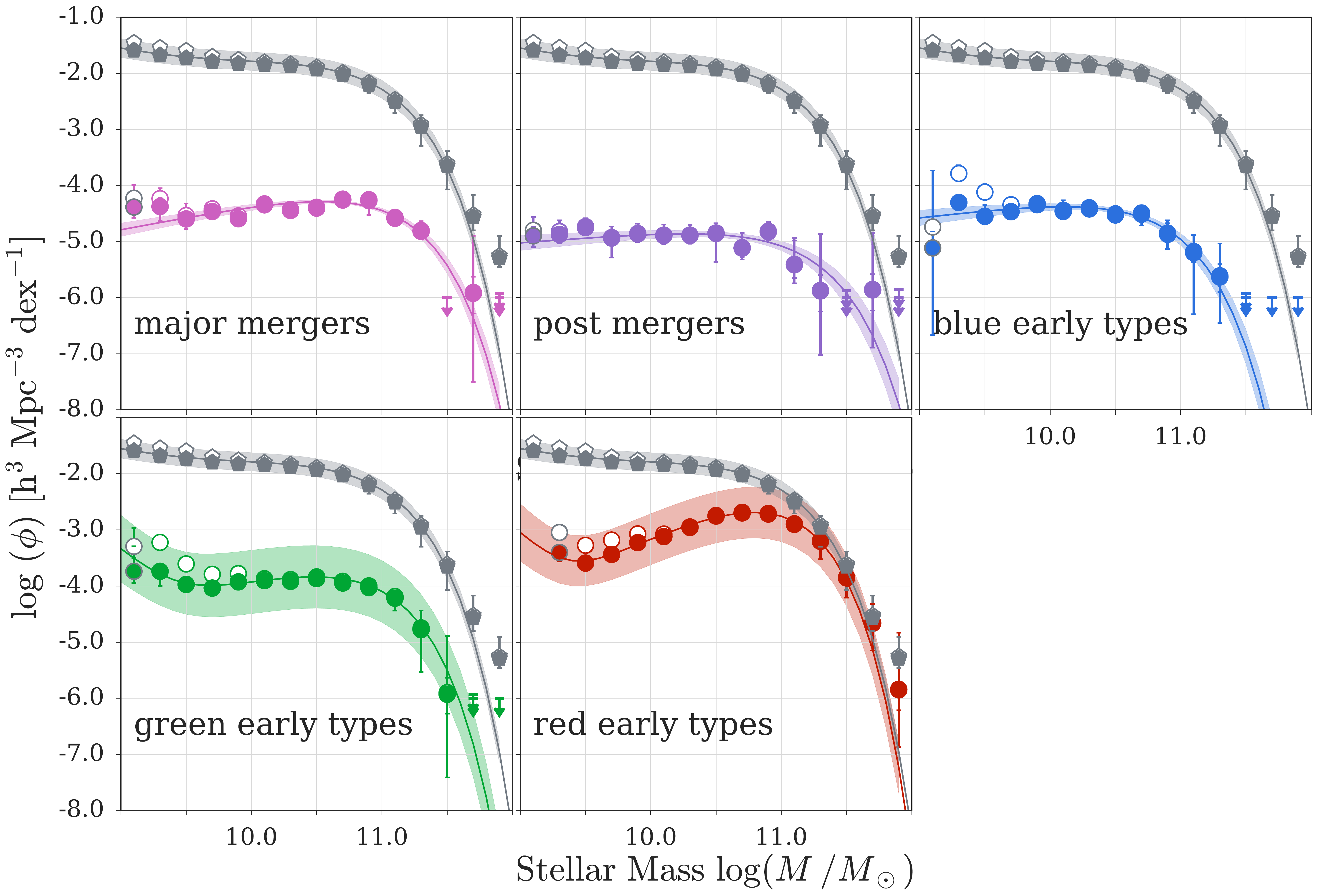}
	\caption{\label{fig:sequence} Stellar mass functions for major mergers, post mergers, blue early types, green early types and red early types. Shown in grey  is the mass function of the entire galaxy sample. Results from the classical 1/\Vmax\ approach \protect\citep{Schmidt:1968aa} and the non-parametric maximum likelihood technique (SWML, \protect\citealt{Efstathiou:1988aa}) are shown with open and filled symbols, respectively.  The best fit Schechter functions according to the parametric maximum likelihood approach (STY, \protect\citealt{Sandage:1979aa}) are illustrated with solid lines. The STY parameters of each subsample are given in Table \ref{tab:parameters}. The shaded region shows the $1\sigma$ uncertainty according to the STY technique. The error bars on the 1/\Vmax\ results correspond to the $1\sigma$ random error. The error bars on the SWML points show the combination of random errors and the systematic error due to stellar mass uncertainties. For stellar mass bins that do not contain any sources we compute upper limits with the 1/\Vmax\ and SWML method. Upper limits are computed for the SWML and 1/\Vmax\ results which is why for some of the mass bins we show two upper limits. See \protect\cite{Weigel:2016aa} for more details.}
\end{figure*}

\begin{figure*}
	\includegraphics[width=\textwidth]{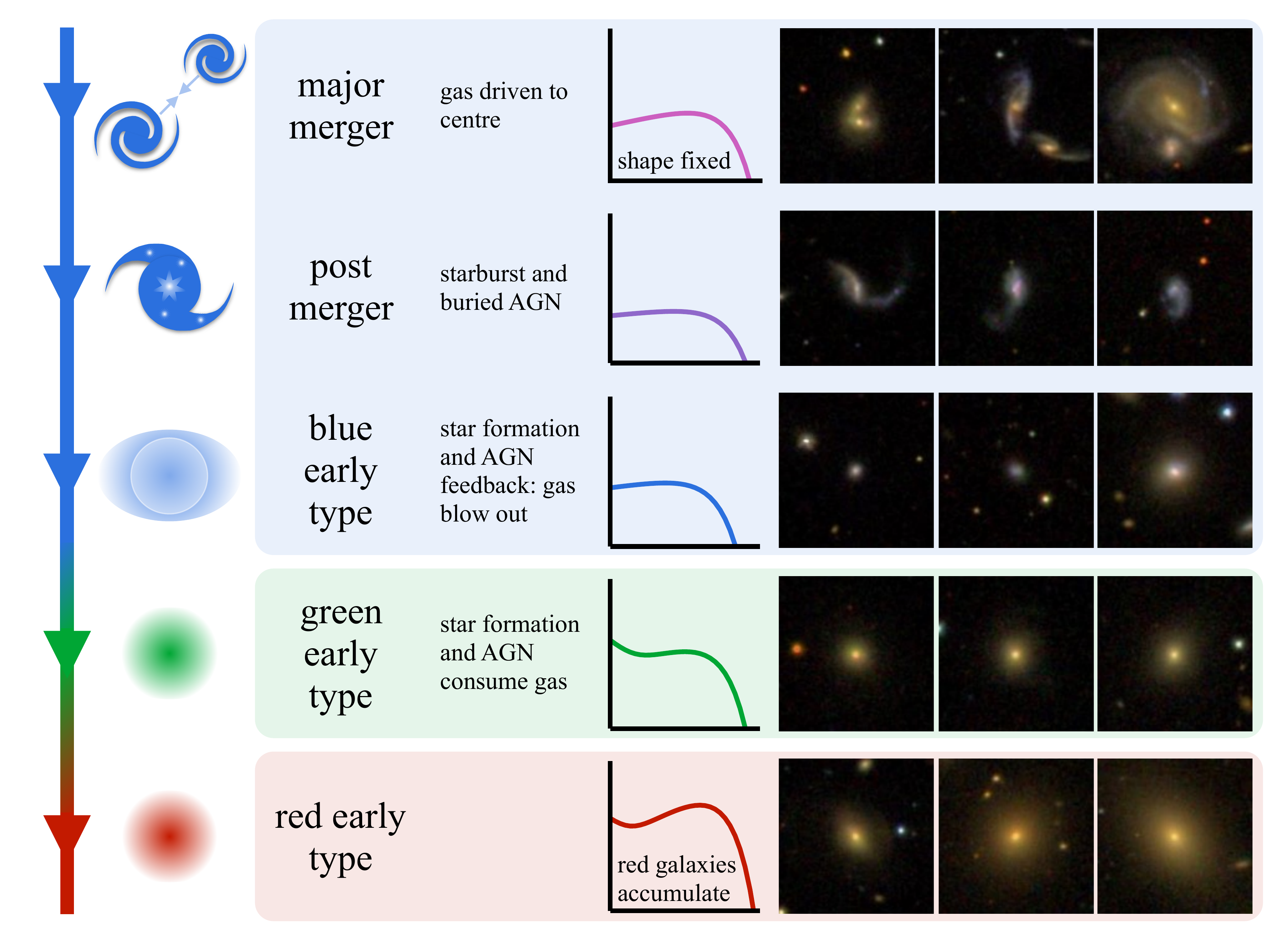}
	\caption{\label{fig:merger_cartoon}Schematic figure illustrating the process of major merger quenching. We highlight the five main stages that we would expect a major merger quenched galaxy to pass through and show the corresponding stellar mass functions and SDSS example images. Note that the stellar mass functions illustrated here are a simplified version of the measured mass functions which we show in Fig. \ref{fig:sequence}. All galaxy samples used here are based on visual classifications from Galaxy Zoo volunteers \protect\citep{Lintott:2008aa,Lintott:2011aa}. The main difference between the major merger and the post merger sample is the number of nuclei: major mergers contain at least two, whereas post mergers only show one  nucleus (\protect\citealt{Carpineti:2012aa} ,\citetalias{Darg:2010aa}, \citealt{Darg:2010ab}). As we discuss in more detail in the text, a merger between galaxies of comparable mass leads to the gas falling towards the center of the merger remnant. This can ignite both a starburst and an AGN. Their feedback can lead to quenching and an evolution of the merger remnant from the blue cloud to the red sequence. We use the stellar mass functions of galaxies along this quenching pathway to study the process of major merger quenching and to estimate its significance.}
\end{figure*}

\begin{figure*}
	\begin{center}
	\includegraphics[width=0.66\textwidth]{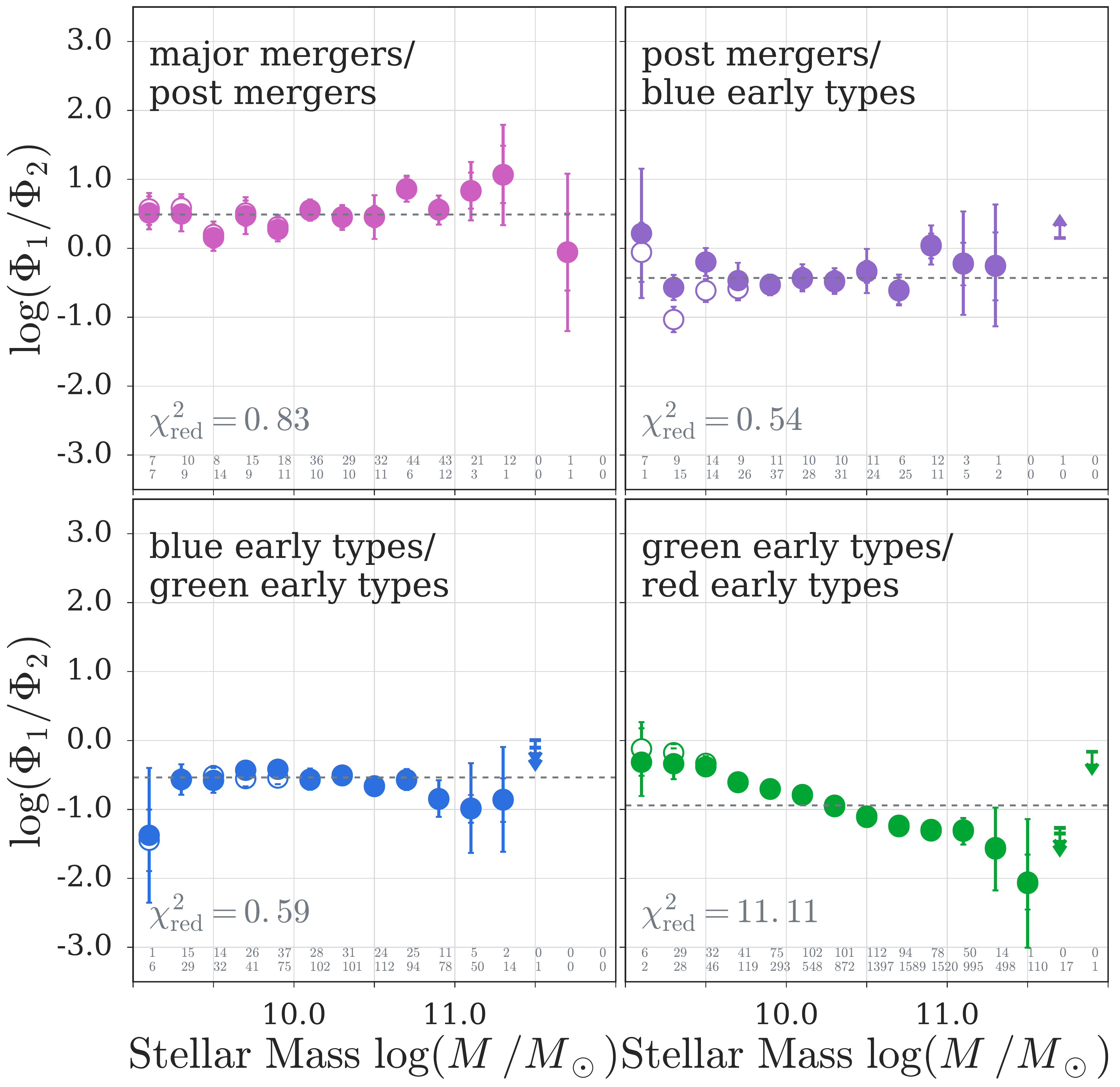}
	\caption{\label{fig:trans_curves} Transition curves for the major merger sequence. We expect the stellar mass functions of galaxies evolving from the major merger towards the red early type stage to have similar shapes if none of the transitions is mass dependent. To test if this is the case and make a possible mass dependence more apparent, we take the ratio between the stellar mass functions of consecutive stages. We refer to these functions as `transition curves' and expect them to be flat if there is no mass dependence. From top left to bottom right we show the transition curves for the major merger to post merger, post merger to blue early type, blue early type to green early type and green early type to red early type stage. Open and filled symbols show the ratio based on the 1/\Vmax\ and SWML results, respectively. To compute upper and lower limits we also use the 1/\Vmax\ and SWML $\Phi$ values which is why in some panels we show two limits for the same mass bin. The horizontal dashed lines show the best fitting relation if we assume a constant fraction. The corresponding $\chi^{2}_{\rm reduced}$ values are given within the panels. Note that upper and lower limits are not included in the fit. We also show the number of objects in each mass bin at the bottom of each panel. The upper and lower rows show the number of galaxies which were used to generate the numerator and denominator stellar mass functions, respectively. This figure illustrates that while the transition curves in the top three panels are consistent with being flat, there is a clear mass dependence between the green and red early type stage (bottom right-hand panel). We explore this trend and its implications more in Section \ref{sec:compare_red}.}
	\end{center}
\end{figure*}

Besides the stellar mass functions of major merger and post mergers, which we introduced in Section \ref{sec:smf_mm_pm}, we also determine the mass functions of blue early types, green early types and red early types. Fig. \ref{fig:sequence} summarizes the stellar mass functions of all five merger quenching sequence stages. For comparison, we also show the stellar mass function of the entire galaxy sample in grey. In analogy to Fig. \ref{fig:merger_post_merger_smf}, open (1/\Vmax) and filled (SWML) symbols and solid lines (STY) show the results of different stellar mass function estimators. For 1/\Vmax\ and SWML we show upper limits in stellar mass bins that do not contain any sources. The best-fitting  STY Schechter function parameters and their errors are given in Table \ref{tab:parameters}. Fig. \ref{fig:merger_cartoon} summarizes and illustrates the major merger quenching sequence.

As we have discussed above, based on our assumptions, we expect the stellar mass functions of galaxies along the quenching sequence to have similar shapes if none of the transitions is mass dependent. To test if this is the case for the mass functions that we are considering here and to make a possible mass dependence more apparent, we take the ratio between mass functions of consecutive steps along the quenching sequence. We refer to these ratios as `transition curves' and illustrate them in Fig. \ref{fig:trans_curves}. 

From top left to bottom right we show the transition curves for the major merger to post merger, the post merger to blue early type, the blue early type to green early type and the green early type to red early type stages. For stellar mass functions of similar shapes these transition curves are flat. Their normalization corresponds to the fraction of galaxies transitioning from one phase to the next, if we assume that the galaxies spend the same amount of time in each stage. 

To quantify the flatness of the transition curves we compute $\chi^2_{\rm reduced}$ values for constant fractions. The $\chi^2_{\rm reduced}$ values are given within Fig. \ref{fig:trans_curves}. Note that for the $\chi^2_{\rm reduced}$ computation we use the SWML data points and we do not take upper and lower limits into account. Based on the $\chi^2_{\rm reduced}$ values, Fig. \ref{fig:trans_curves} shows that the transition curves of most stages are consistent with being flat, only the evolution from the green to the red early type stage shows a significant mass dependence. 

According to the STY method and the likelihood ratio test, the green early type and the red early type mass functions are well described by double Schechter functions. Compared to the green early types, the red early types are however fit by a stronger double Schechter with higher $\log(\Phi^{*}_2/\Phi^{*}_1)$ and $\alpha_2$ values (see Fig. \ref{fig:sequence} and Table \ref{tab:parameters}). The red early types thus have a higher number density at high stellar masses which causes the strong mass dependence that we see in the bottom left panel of Fig. \ref{fig:trans_curves}. We discuss the implications of the green and the red early types having significantly different mass functions in more detail in Section \ref{sec:compare_red}.

We conclude that except for the evolution of green to red early types, the transition curves are consistent with being flat. We thus infer that the galaxies that we find in these different phases today are likely to evolve along a sequence. This sequence bridges from the blue cloud to the red sequence and does not include a significant mass dependence. Hence, major mergers are likely to lead to quenching for a majority of galaxies that are involved in a gas rich merging event.

Our expectation of similar mass function shapes for galaxies along the merger quenching sequence is based on the assumption of a mass independent merger-to-red-early-type transition probability. As we argued in Section \ref{sec:mass_dep_prob}, we assume that the effects of disc reforming, AGN feedback, dynamical friction and fading of merger features do not introduce a significant mass dependence in the probability of a merger remnant reaching the red sequence. Implicitly we also assume that the stellar mass measurements of major mergers and post mergers and their visual classifications are unbiased. The transition curves of galaxies along the merger quenching sequence being flat thus either implies that our assumptions are justified or that two or multiple effects compensate their respective mass dependence.

If there are, for instance, two mass dependent mechanisms which have an opposite effect on the space density of merger quenched galaxies, than they have to both affect the \textit{same} stages along the merger quenching sequence. For example, if low mass galaxies have a high probability of reforming a disc, we would expect these galaxies to be part of the major merger, but maybe not the green early type sample. The space density of low mass green early types would be lower, causing the green early type mass function to have a different slope $\alpha$ than the major merger mass function. This effect could not be compensated by a mass dependence of dynamical friction which would primarily affect major merger and post merger galaxies. Similarly, a systematic bias could affect the stellar mass measurements of  major merger and post merger galaxies. Yet we do not expect the stellar mass measurements of standard ellipticals to be affected by biases. The blue early type mass function being similar in shape to the major merger and post merger mass functions hence provides evidence against a bias in the $M$ measurement of major mergers and post mergers, if we assume that the aforementioned effects are indeed mass independent.

We conclude that the probability of a mass dependence of the effects considered here causing mass functions of similar shapes for all stages along the merger quenching sequence is low.  The simpler and more straightforward explanation for the flat transition curves shown in Fig. \ref{fig:trans_curves} is that these stages do indeed represent an evolutionary sequence and that the previously mentioned effects do not significantly impact the mass distribution of galaxies following this evolution.

\subsection{Transition time scales}
\label{sec:timescales}
\begin{figure}
	\includegraphics[width=\columnwidth]{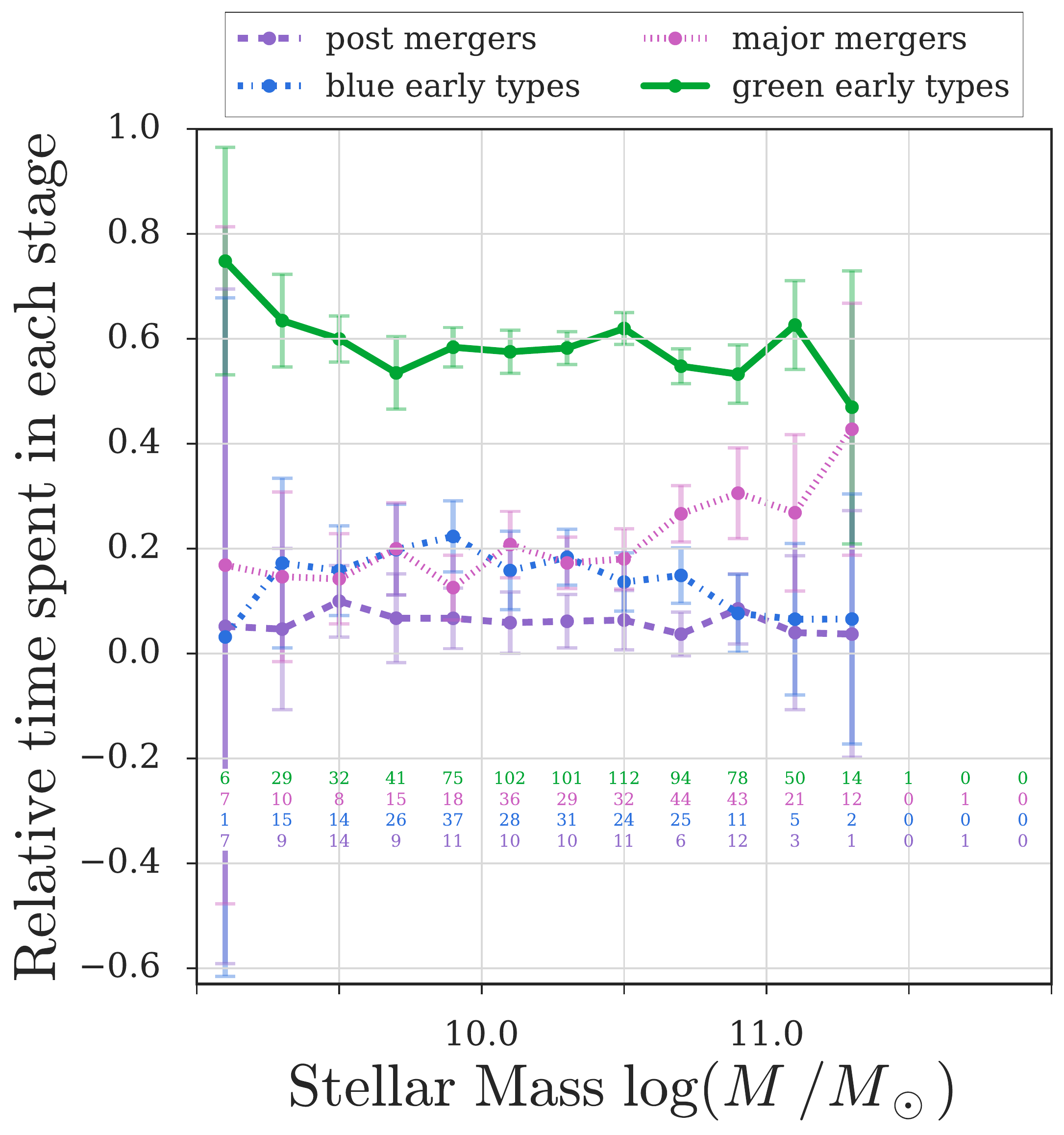}
	\caption{\label{fig:timescales} Fraction of time spent in stages along the major merger sequence relative to the total time it takes to transition from the major merger to the green early type phase. We assume that the majority of galaxies transitions from one stage to the next. Under this assumption, the number density of galaxies in a certain phase is proportional to the time spent in this phase. We use the stellar mass functions to compute the relative amount of time that galaxies spend in the major merger, the post merger, the blue early type and the green early type stage. In the bottom part of the figure we give the number of objects that we considered when computing the stellar mass functions and time scales. From top to bottom we show the number of green early types, major mergers, blue early types and post mergers in each mass bin. Due to the low number density of high mass galaxies, mass bins above $\log (M/M_\odot) = 11.3$ contain very few objects. Thus we do not constrain the relative time scales for these bins.}  
\end{figure}

By assuming that along the major merger sequence most galaxies transition from one phase to the next, we can constrain the relative amount of time spent in each phase. We consider the number density of galaxies in the major merger, post merger, blue early type and green early type stage. We do not take galaxies in the red early type phase into account since all quenched galaxies accumulate in this stage. We use the SWML results and calculate the relative amount of time spend in phase $i$ for mass bin $k$ in the following way:

\begin{equation}
t_i (M_k) = \frac{\Phi_i(M_k)}{\sum_{j}^{N_\mathrm{states}} \Phi_j(M_k)}.
\end{equation}

The $1\sigma$ error on $t_i$ is given by:

\begin{equation}
\small
\begin{aligned}
\sigma_{t_i}^2 &= \sum_{j}^{N_\mathrm{states}} \left(\frac{\partial (\Phi_i/\sum \Phi)}{\partial \Phi_j} \right)^2 \sigma_{\Phi_j}^2\\
& = \frac{1}{\left(\sum \Phi\right)^2} \left(\sum_{j}^{N_\mathrm{states}} \left (\frac{\Phi_j^2 \sigma_{\Phi_j}^2}{(\sum \Phi)^2}\right) + \left(1 - 2 \frac{\Phi_i}{\sum \Phi}\right) \sigma_{\Phi_i}^2\right)
\end{aligned}
\end{equation}

Fig. \ref{fig:timescales} summarises our results and shows that at a given stellar mass galaxies spend $\sim 60\%$ of their transition time in the green early type stage. The post merger sample contains only very few galaxies. This implies that galaxies spend only $\sim5\%$ of their time in the post merger stage. 

The numbers at the bottom of Fig. \ref{fig:timescales} show the number of objects in each mass bin which were used to compute the stellar mass functions of the samples used here (from top to bottom: green early types, major mergers, blue early types, post mergers). Due to the low number density of high mass galaxies, our sample contains only very few galaxies at masses above $\log (M/M_\odot) = 11.3$. For example, the $\log (M/M_\odot) = 11.5$ mass bin only contains one green early type and no major merger, blue early type or post merger galaxy. We thus do not compute $t_i$ for $\log (M/M_\odot) > 11.3$.

\subsection{Comparing the mass functions of major mergers and red early type galaxies}
\label{sec:compare_red}
\begin{figure}
	\includegraphics[width=\columnwidth]{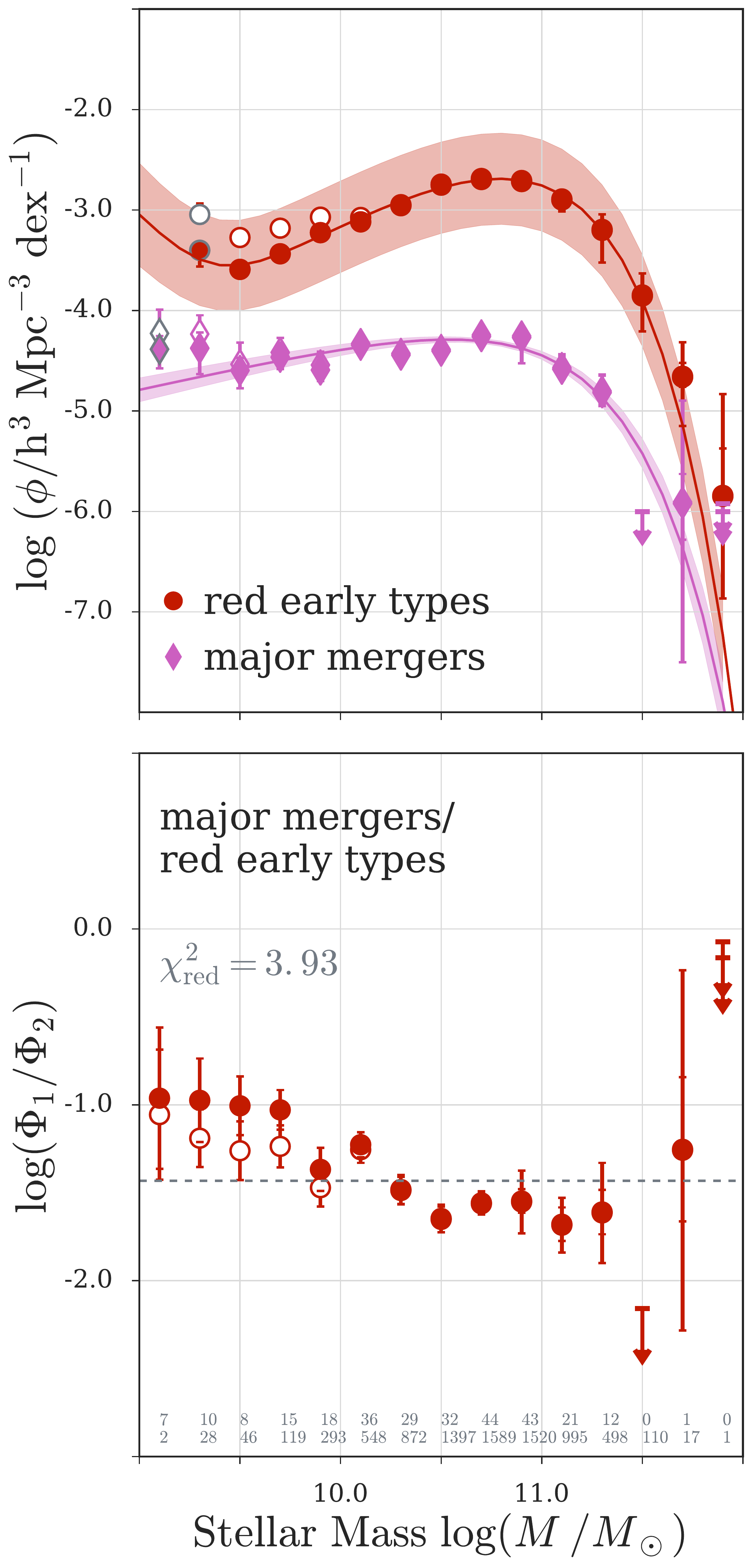}
	\caption{\label{fig:comparison_red_red_ET_mm}Stellar mass functions of major mergers and red early type galaxies and the corresponding transition curve. The upper panel shows that major mergers and red early type galaxies have significantly different stellar mass function shapes. This can also be seen in the bottom panel which shows the transition curve for these two subsamples.  As we discuss in the text, the significantly different shapes of the red early type and the major merger mass function imply that it is unlikely that all red early types have been quenched through the merger quenching process we observe in the local Universe. Other quenching processes must either lead to the formation of red early type galaxies, or the process of merger quenching and thus the major merger mass function must have evolved with time. In analogy to to previous figures, we show the best fitting constant fraction and the corresponding $\chi^2_{\rm reduced}$ value in the bottom panel. The number of objects in each mass bin are given at the bottom of the panel (numerator sample at the top, denominator sample at the bottom). In both panels, 1/\Vmax\ and SWML results are shown with open and filled symbols, respectively. Upper limits are compute based on both methods.}
\end{figure}

We expect the red early type stage to be the final phase in the evolution of major merger quenched galaxies. By comparing the shapes of the red early type mass function and the major merger mass function we can make inferences about the physical mechanisms that might be building up the red sequence. 

We make the following statement:
\begin{itemize}
	\item if all red early type galaxies have been quenched through major mergers,
	\item and if red early type galaxies do not gain significant amounts of mass while on the red sequence, 
\end{itemize}
the mass function of red early type galaxies should resemble the mass function of major mergers. 

As discussed in Section \ref{sec:mm_quenching}, we expect galaxies evolving along the major merger quenching sequence to not increase their stellar mass significantly, thereby maintaining the same mass distribution. On the red sequence, major merger quenched galaxies accumulate. This leads to an increase in their number density over time. The shape of their stellar mass function however stays constant, if they do not gain significant amounts of mass. The red early type mass function should thus be as flat as the mass functions of previous stages along the merger quenching sequence, if the aforementioned assumptions are true.   

When discussing the transition curves in Section \ref{sec:mm_mf}, we already pointed out that the green and red early type mass functions have significantly different shapes. In Fig. \ref{fig:comparison_red_red_ET_mm} we compare the major merger mass function to the mass functions of red early type galaxies. 

The red early type mass function has a  significantly different shape compared to the major merger mass function. While the major merger mass function is fit with a flat single Schechter function, the red early type mass functions has a strong double Schechter form. 

This contradicts our expectations outlined above and implies that:
\begin{itemize}
	\item the mass function of red early type galaxies has evolved with time, 
	\item or that the process of major merger quenching and thus the major merger mass function have evolved with time,
	\item or that red early type galaxies can be created through alternative physical process, not just major merger quenching. 
\end{itemize} 

The only way a red galaxy can increase its stellar mass significantly is through a dry merger. \citetalias{Peng:2010aa} discuss the effect that dry merging might have on the red stellar mass function shape.

They argue that dry mergers can primarily affect the steep high mass end of the stellar mass function. To investigate the change in the red mass function they use a single Schechter function with $\alpha = -1.4$ and assume that $15\%$ of all red galaxies undergo a 1:1 merger. Note that their dry merger probability is mass independent. After the merging the new population of red galaxies consists of two groups: the $85\%$ of red galaxies that did not undergo a major merger and retained their stellar mass function shape and the population of merged galaxies which has increased its $M^{*}$ by 0.3 dex. Due to the major mergers the new population of red galaxies contains fewer galaxies and has increased its stellar mass by 0.03 dex on average. When fitting the combined population of merged and unmerged galaxies with a single Schechter function, \citetalias{Peng:2010aa} find an $M^{*}$ increase of 0.09 dex and a steepening of $\alpha$ by 0.15. The fact that $M^{*}$ increased by 0.09 and not 0.03 dex is due to the degeneracy between $\alpha$ and $M^{*}$ in the single Schechter fitting. Mergers with higher mass ratios will results in smaller changes in $M^{*}$. The simple model by \citetalias{Peng:2010aa} thus shows that dry mergers might lead to an increase in $M^{*}$ and a steepening of $\alpha$. 

Based on the results by \citetalias{Peng:2010aa}, we conclude that dry merging is unlikely to be the cause of the significant difference between the flat mass functions of the merging sequence and the red early type mass function. For example, if the red early type mass function had an initial shape similar to the green early type mass function, dry merging would cause $\alpha_2$ to decrease. Yet we observe $\alpha_{2, \rm red} > \alpha_{2, \rm green}$. An evolution in the red early type mass function due to dry merging that could explain the discrepancy between the major merger and the red early type mass function which we observe at $z\sim0$ is thus less likely.

A change in the major merger mass function shape with time is possible. At higher redshift major merger quenching could for instance only lead to early type formation at high stellar masses. Lower mass galaxies could be gas-rich enough to reform a disc and would thus not be part of the red early type sample \citep{De-Lucia:2011aa, Rodriguez-Gomez:2016aa}. This could lead to a build up of red early type galaxies at high stellar masses, thus explaining the strong double Schechter function shape that we observe at $z\sim0$. Repeating the analysis at higher redshift would allow us to test this possibility. In the local Universe the reforming of a disc at low stellar masses is unlikely to be a significant effect. Low mass galaxies that reform a disc would be part of the major merger, but for instance not the blue early type sample. This would cause major mergers and galaxies in subsequent stages to have different mass function shapes, which is inconsistent with the observations (see Section \ref{sec:mm_mf}). 

The red early type mass function shape could also be explained by the existence of alternative quenching channels. Major merger quenching explains the existence of elliptical galaxies in the green valley. The green valley transition zone does however also include late types and indeterminates, which are galaxies that, based on their vote fraction distribution, can neither be classified as clear early nor as clear late types (see Section \ref{sec:galaxy_sample}). Late types and indeterminates are less likely to have been quenched through major mergers and could, through changes in their morphology, become part of the red early type population.

The shape of the red early type mass function therefore implies that it is unlikely that all red early type galaxies have been created through the major merger quenching process that we observe today.  Alternative quenching channels must lead to red early type formation or merger quenching at higher $z$ must have led to a different mass function shape.  
\subsection{Comparing the mass functions of major mergers and green galaxies}\label{sec:compare_green}
\begin{figure}
	\includegraphics[width=\columnwidth]{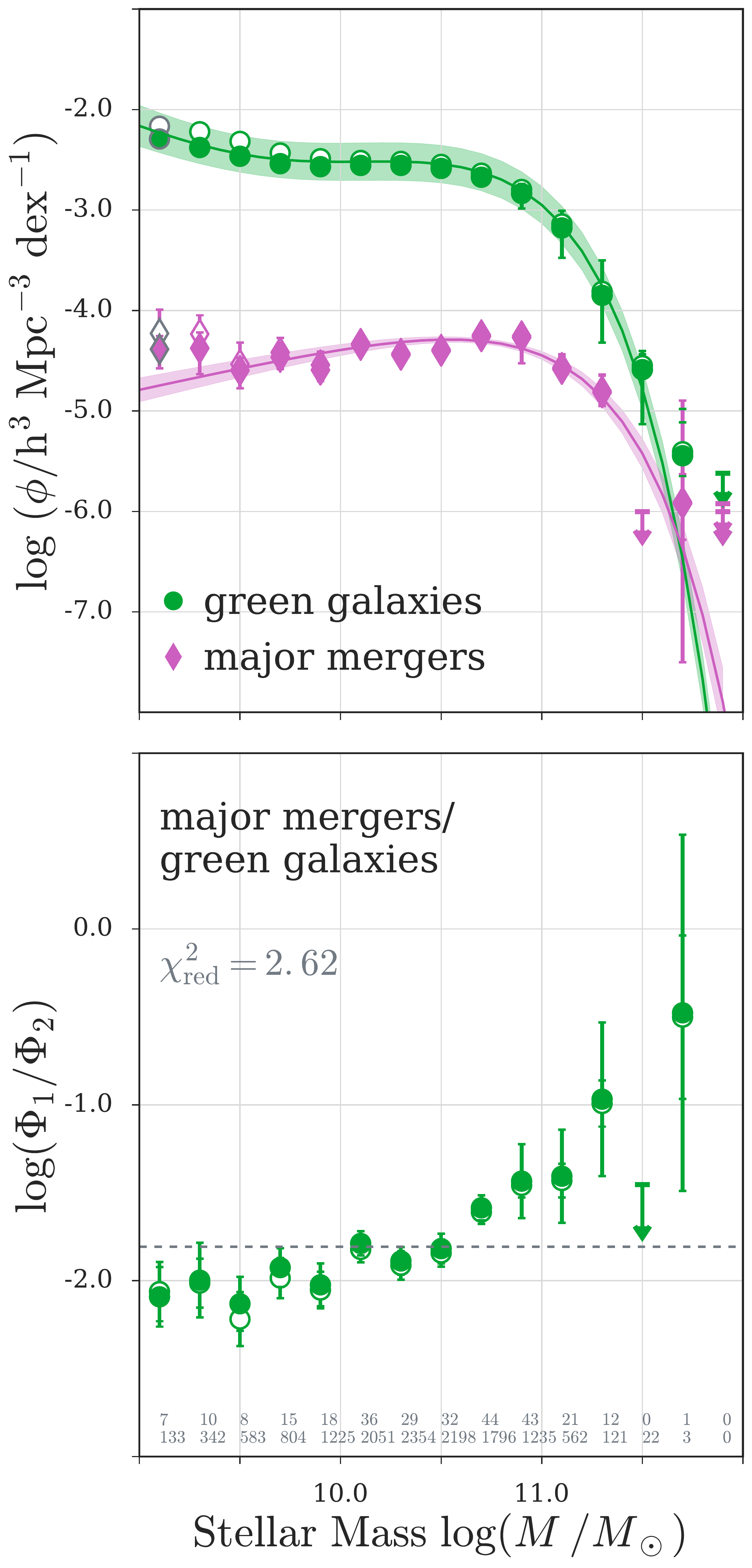}
	\caption{\label{fig:comp_mm_green_mf}Stellar mass functions of major mergers and green galaxies and the corresponding transition curve. The figure illustrates that major mergers and green galaxies have significantly different mass function shapes. This implies that not all green galaxies have been major merger quenched and that the green valley is likely to be dominated by a population of galaxies that have been quenched through alternative quenching channels.This argument is only based on the mass functions presented here and does not depend on morphological classifications. For the mass function ratio we indicate the best fitting constant fraction with a dashed line and show the corresponding $\chi^2_{\rm reduced}$ value. The numbers in the bottom panel show the number of objects per mass bin (upper row: major mergers, bottom row: green galaxies). 1/\Vmax\ and SWML results are shown with open and filled symbols, respectively. Upper limits are computed based on both methods.}
\end{figure}

In analogy to the previous section, we compare the stellar mass function shapes of major mergers and green galaxies. Fig. \ref{fig:comp_mm_green_mf} shows the stellar mass functions of major mergers and green galaxies in the top panel and the ratio of these functions in the bottom panel. 

Based on our assumptions, we expect the mass function of green galaxies that are merger remnants to be similar in shape to the major merger mass function. If the majority of green galaxies was merger quenched, the mass function of green galaxies should also resemble the major merger mass function. The bottom panel of Fig. \ref{fig:comp_mm_green_mf} shows that the green and the major merger mass functions have significantly different shapes. The green galaxy population does therefore not only consist of major merger quenched galaxies and must be dominated by a population of galaxies that have been quenched through alternative quenching channels. 

Note that this argument is only based on the shapes of the major merger and the green galaxy mass functions and the assumptions that merger quenched galaxies reach the green valley at some point and do not gain significant amounts of mass during their evolution. The statement is independent of the mass functions of post mergers, blue early types and green early types and does therefore not depend on morphological classifications.  

\subsection{Merger contribution to the flux through the green valley}\label{sec:green_valley_flux}
\begin{figure*}
	\includegraphics[width=\textwidth]{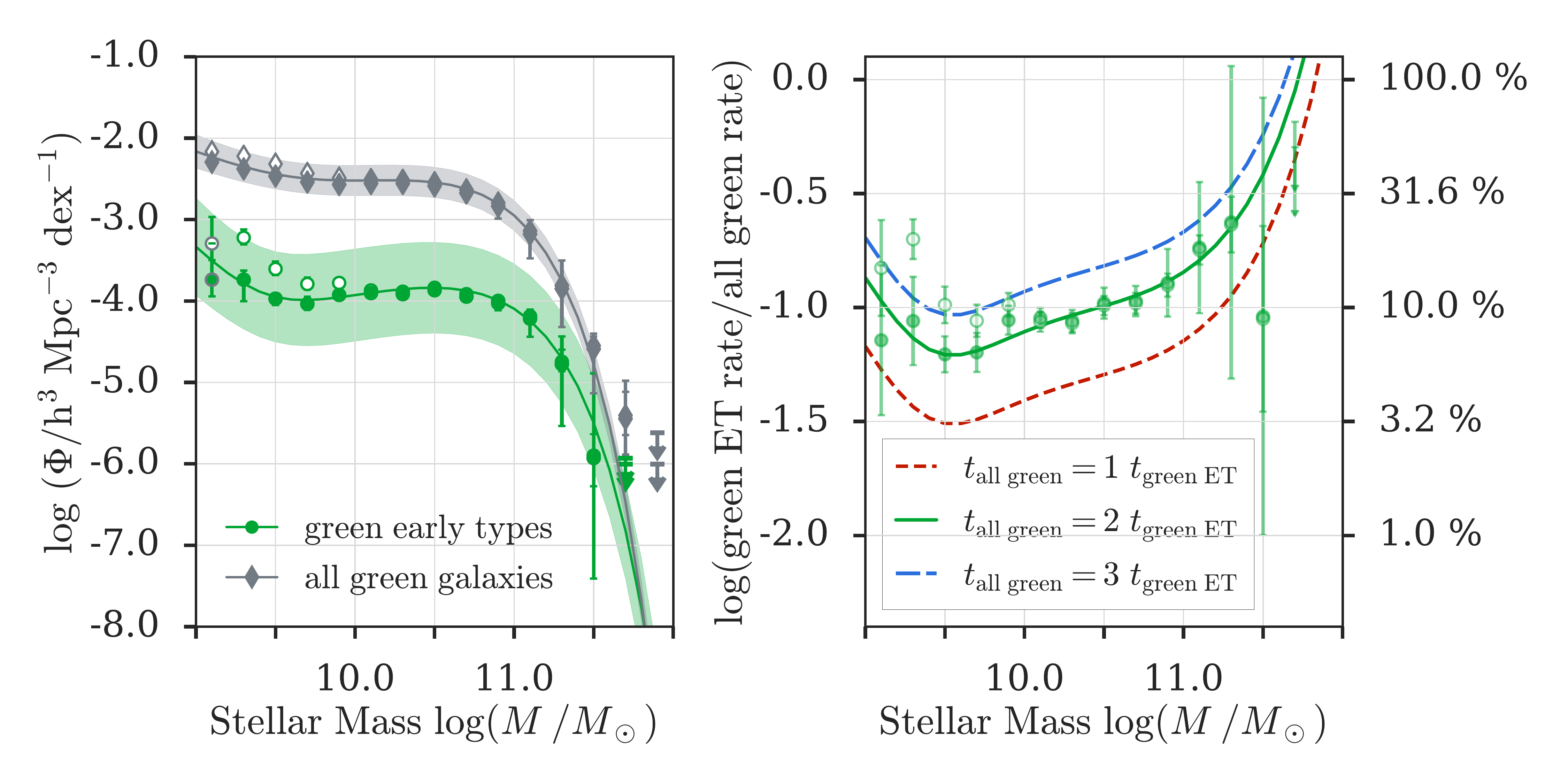}
	\caption{\label{fig:green_valley_frac}Major merger contribution to quenching at $z\sim0$. To estimate the contribution of major merger quenching to the quenching of galaxies in the local Universe, we use the number density of early types in the green valley. These galaxies are likely to have experienced a major merger in the past and will reach the red sequence in the future. In the left-hand panel we show the stellar mass functions of all green and of green early type galaxies. The number density of green early types is significantly lower than the number density of all green galaxies. For instance, the difference in $\Phi^{*}$ is $\sim 1.3$ dex corresponding to $\sim 5\%$. From this we could already conclude that major merger quenched galaxies only make up a small fraction of all galaxies that are quenching at $z\sim0$. However \protect\cite{Schawinski:2014aa} and \protect\cite{Smethurst:2015aa} have shown that the green valley transition time is morphology dependent. Green early type galaxies tend to transition the green valley on shorter time scales than late type galaxies. This increases the merger quenching contribution to the green valley flux and quenching at $z\sim0$. In the right-hand panel we show the green early type transition rate relative to the transition rate of all green galaxies. For the short dashed, solid and long dashed lines we assumed that on average green early types cross the green valley as fast as, two times as fast as and three times as fast as all green galaxies. These ratios are based on the STY results. For clarity we show the ratio based on the 1/\Vmax\ (open symbols) and SWML (filled symbols) results for $t_{\rm all\ green} = 2 t_{\rm green\ ET}$ only. The precise value of the flux ratio depends on, for example, the definition of the green valley and the vote fraction threshold used for the morphological classifications. Nonetheless, this figure illustrates that major merger quenched galaxies are unlikely to make up the majority of galaxies transitioning the green valley at $z\sim0$. Due to their low number density this holds even if we take morphology dependent green valley transition times into account.}
\end{figure*}
After the very general arguments in the previous sections, we now use the green early type mass function to discuss the significance of merger quenching at $z\sim0$ in more detail. In Section \ref{sec:red_sequence} we model the redshift evolution of the red sequence and determine the fraction of galaxies that have been quenched through major mergers. 

Galaxies which lie in the green valley at $z\sim0$ are currently transitioning from the blue cloud to the red sequence. They have experienced a physical process which initiated their change in colour in the past and they will be reaching the red sequence in the future. In Section \ref{sec:theory} we discussed why major merger quenched  galaxies appear to already have an early type morphology once they enter the green valley. In the following discussion we will assume that the majority of green early type galaxies has been quenched through major mergers. Morphologies then allow us to identify merger quenched galaxies and to estimate their contribution to quenching at $z\sim0$. 

In the left-hand panel of Fig. \ref{fig:green_valley_frac} we show the stellar mass functions of green early type galaxies and of all green galaxies. The figure illustrates that green early type galaxies only make up a small fraction of the green valley population. The difference in $\Phi^{*}$, for instance, is $\sim 1.3$ dex, corresponding to $\sim 5\%$.  

\cite{Schawinski:2014aa} use the NUV-optical colour-colour diagram to investigate the evolution of early and late type galaxies across the green valley. They show that early type galaxies transition the green valley on a time scale on the order of 1 Gyr, whereas late type galaxies evolve on a time scale on the order of several Gyr. Using a Bayesian approach, \cite{Smethurst:2015aa} use $\tau$-models to constrain the star formation history of individual galaxies. They  measure a range of quenching time scales for smooth- and disc-like galaxies. In the green valley and at $z\lesssim2$ the majority of bulge and disc dominated galaxies quench on intermediate ($1 < \tau/\rm Gyr < 2$) and slow ($\tau/\rm Gyr > 2$) time scales, respectively. Rapid quenching time scales ($\tau/\rm Gyr < 1$) are more likely to be found for smooth-like than disc-like galaxies.

Green early type galaxies make up a small fraction of the total \textit{number} of green galaxies. Yet their contribution to the overall number \textit{flux} of galaxies across the green valley increases, if on average they evolve faster than the entire green valley population.

We use stellar mass functions to compare the number flux of green early type galaxies to the number flux of all green galaxies. We weigh the stellar mass functions of green early types and of all green galaxies with the corresponding green valley transition times ($t_{\rm green\ ET}$, $t_{\rm all\ green}$) and determine the ratio of these rates. Note that the transition times we use here do not correspond to the $\tau$-model quenching time scale, i.e. they do not describe the exponential decline of the SFR. Instead $t_{\rm green\ ET}$ and $t_{\rm all\ green}$ correspond to the average green valley transition time of green early type and all green galaxies. The contribution of green early types to the overall green valley flux is proportional to the ratio of $t_{\rm green\ ET}$ to $t_{\rm all\ green}$ and not their absolute values. 

The right-hand panel of Fig. \ref{fig:green_valley_frac} illustrates our results. We show the ratio of the green early type rate to the overall green rate for $t_{\rm all\ green} /t_{\rm green\ ET} = 1, 2, 3$. For instance, using the result by \cite{Schawinski:2007aa} and \cite{Schawinski:2014aa} and assuming $t_{\rm green\ ET} \sim 1$ Gyr this implies that all green galaxies transition the green valley within $1 - 3$ Gyr. Note that a factor of three difference in the green valley transition times can correspond to an orders of magnitude difference in terms of $\tau$. The lines show the ratio based on the STY results. For clarity we show the results based on the 1/\Vmax\ and SWML for $t_{\rm all\ green} /t_{\rm green\ ET} = 2$ only. The green galaxies sample contains few objects at high stellar masses. This is reflected by upper limits and large error bars at the high mass end of the green and the green early type mass functions. The high mass end of the flux ratio is thus affected by large uncertainties. The upturn at $\log M \gtrsim 11$ in the right-hand panel of Fig. \ref{fig:green_valley_frac} should hence not be taken at face value. Fig. \ref{fig:green_valley_frac} shows that at a given stellar mass between $9 < \log(M/M_\odot) < 11$ green early type galaxies make up $\sim 3-30\%$ of the the overall green valley flux if $1 < t_{\rm all\ green} /t_{\rm green\ ET} < 3$.

We stress that the fraction of $\sim 3-30\%$ represents a zeroth-order estimate. Besides being proportional to $t_{\rm all\ green} $ and $t_{\rm green\ ET}$, the fraction does, for instance, depend on the distribution of quenching times, i.e. the times at which galaxies start their evolution from the blue cloud to the red sequence. Furthermore, the definition of the green valley and the green early type sample affect the results. As we have discussed in Section \ref{sec:galaxy_sample}, the morphological classifications we use are based on the vote fractions of Galaxy Zoo users. For example, lowering the vote fraction cut above which we define a galaxy as an early type would result in some of the galaxies that are currently part of the indeterminate category to be assigned to the early type sample. The number of green early type galaxies and thus their contribution to the green valley flux would increase. We also assume that the majority of green early type galaxies at $z\sim0$ have been quenched through major mergers. If alternative quenching processes lead to the formation of green early type galaxies, assuming that major merger quenched galaxies account for $\sim 3-30\%$ of the green valley flux would be an overestimate. 

The analysis presented in this section does however show that it is unlikely that the majority of galaxies that are transitioning the green valley at $z\sim0$ have been quenched through major mergers. This is due to the low number of green early type galaxies and holds even if we take into account that the green valley transition time does depend on morphology.

\subsection{Merger contribution to the build up of the red sequence}\label{sec:red_sequence}

\begin{figure*}
	\includegraphics[width=\textwidth]{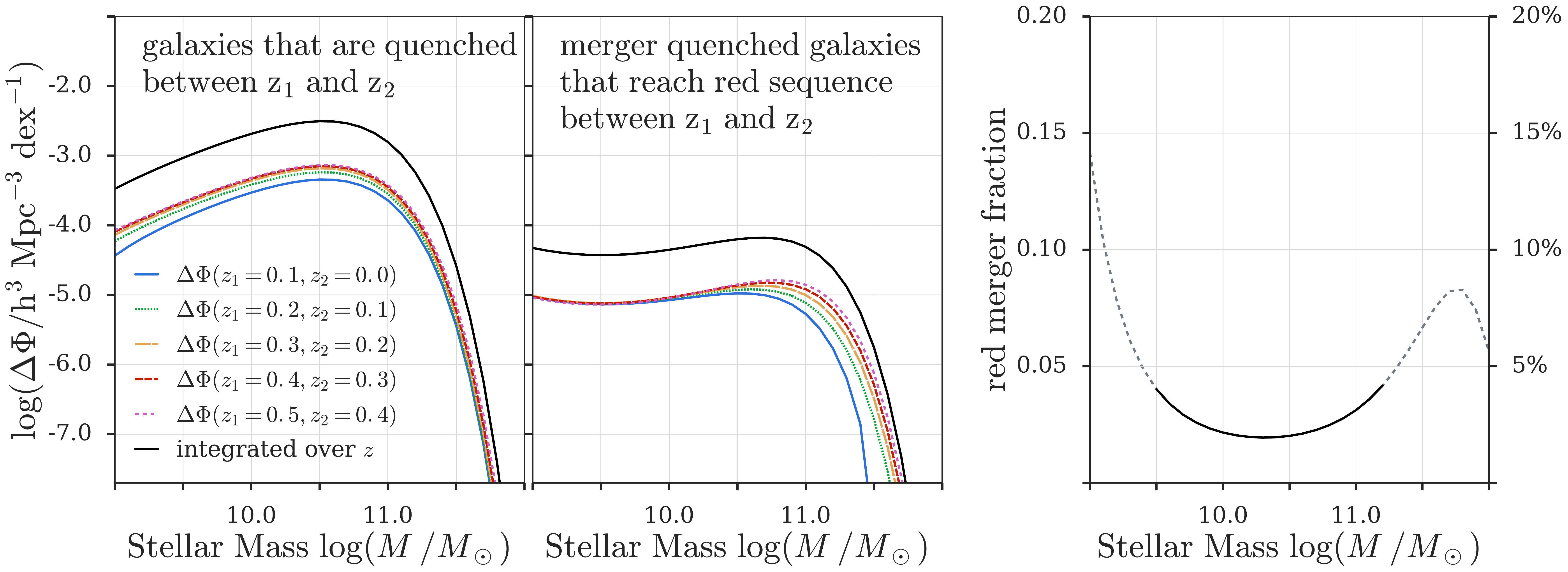}
	\caption{\label{fig:red_blue_toy_model}Contribution of merger quenched galaxies to the build up of the red sequence between $0 \leq z \leq 0.5$. We model the evolution of the red (left-hand panel) and the major merger mass function (middle panel) to determine what fraction of red galaxies was quenched through major mergers (right panel). To estimate the number density of mass quenched galaxies between $z_1$ and $z_2$ ($z_1 > z_2$), we predict the shape of the blue mass function at $z_2$ by evolving the blue galaxies along the main sequence and then subtracting the true, blue mass function at $z_2$.  We model the redshift evolution of the major merger mass function by combining our measurement of the merger fraction at $z \sim 0$ (see Fig. \ref{fig:fraction}) with the redshift evolution of the blue mass function \protect\citep{Caplar:2015aa} and a redshift dependent normalization to allow for more merging at higher $z$ \protect\citep{Bridge:2010aa}. We then derive the mass function of merger quenched galaxies by assuming that it takes a galaxies about 1 Gyr to transition form the major merger to the red early type stage. We integrate over $z$ and measure the contribution of merger quenching to the red sequence by taking the ratio of all merger quenched and all mass quenched galaxies. We show the red merger fraction as a function of stellar mass in the right-hand panel. The dashed line shows the masses at which the major merger (low mass and high mass end) and the red stellar mass function (high mass end) are affected by significant uncertainties.}
\end{figure*}

To estimate how much quenching through major mergers could have contributed to the development of the red sequence we model the evolution of the major merger and the red galaxy mass function over the redshift range $0 \leq z \leq 0.5$. 

In addition to the points outlined in Section \ref{sec:assumptions}, we make the following assumptions:
\begin{enumerate}
	\item{Most galaxies that were involved in a major merger transition to the red sequence (see our discussion in Section \ref{sec:mass_dep_prob}).}
	\item{It takes galaxies of the order of $t_{\rm mm} \sim 1$ Gyr to transition from the major merger to the red early type stage (see e.g. \citealt{Springel:2005ad}).}
	\item{The major merger mass function retains it $z\sim0$ shape and does not change with $z$.}
	\item{The normalization of the major merger mass function increases as $(1 + z_\mathrm{mm})^\beta$. Based on \cite{Bridge:2010aa}, we use $\beta = 2.25$. }
\end{enumerate}

So far, previous studies have primarily studied the redshift evolution of the  integrated merger fraction (see e.g. \citealt{Bridge:2010aa, Lotz:2011aa, Robotham:2014aa, Keenan:2014aa}). Studying how the merger mass function, or equivalently the merger fraction as a function of stellar mass, evolves with $z$ is more challenging. We assume that the merger mass function retains its shape and that only $\Phi^{*}$ increases with increasing $z$. This represents the simplest and most straightforward assumption that we can make until the $z$ evolution of the merger mass function has been constrained observationally.

First, we motivate our assumption for $t_{\rm mm}$. Second, we estimate the number density of major merger quenched galaxies reaching the red sequence within a $z$ interval. Third, we find an analytic expression for the number density of newly quenched galaxies within a redshift interval. By comparing the two quantities and integrating over $z$, we determine the contribution of major merger quenched galaxies to the build up of the red sequence.

\subsubsection{Time scales}
Observationally it is challenging to determine the average time that it takes a galaxy to evolve from the major merger to the red early type stage. Besides visual classification, common methods to identify major mergers include the close pair technique (see e.g. \citealt{Patton:2000aa, Ellison:2008aa}) and morphological measurements based on CAS (see e.g. \citealt{Conselice:2003ab}), the Gini coefficient or $M_{20}$ (see e.g. \citealt{Lotz:2004aa}). While these techniques allow the measurement of the merger fraction, estimating the merger rate is less straight forward and can lead to disagreement among the methods  \citep{Lotz:2011aa}. For the CAS method \cite{Bertone:2009aa} use a sensitivity time scale of 0.4 to 1 Gyr and for the close pair technique \cite{Patton:1997aa} assume an infall time of the order of a few hundred Million years. 

Using simulations,  both \cite{Springel:2005ad} and \cite{Hopkins:2008ab} find that merging galaxies significantly change their optical colour and decrease their SFR within less than 1 Gyr if AGN feedback is considered. Also invoking AGN feedback, \cite{Kaviraj:2011aa} use a phenomenological model to reproduce the $\sim$ 1 Gyr long blue-to-red evolution of elliptical galaxies which was previously observed by  \cite{Schawinski:2007aa}.

For our simple model we thus assume $t_{\rm mm} \sim$ 1 Gyr for the average transition time between the major merger and the red early type stage. In Fig. \ref{fig:diff_timescale} we furthermore show that the chosen $t_{\rm mm}$ value does not significantly affect our results by varying $t_{\rm mm}$ from 0.5 to 3 Gyr.

\subsubsection{The evolution of the major merger mass function}

In the top middle panel of Fig. \ref{fig:fraction} we show the ratio between the major merger mass function and the mass function of all blue galaxies. We can interpret this ratio as the merger fraction relative to all blue galaxies and rewrite the major merger mass function as
\begin{equation}
\Phi_\mathrm{mergers}(z=0, M) = \mathrm{frac}(M) \Phi_\mathrm{blue}(z=0, M)
\end{equation}
We determine the mass function of galaxies that were quenched through a major merger and are reaching the red sequence at redshift $z$ by considering the blue galaxy mass function at the merging time, $z_\mathrm{mm}$ ($z_\mathrm{mm} - z  \mathrel{\hat{=}} t_{\rm mm} \sim 1$\ Gyr). The mass function of galaxies that are reaching the red sequence at $z$ and were involved in a major merger at $z_\mathrm{mm}$ can thus be expressed as
\begin{equation}\label{eq:mergers_z}
\Phi_\mathrm{red\ mergers}(z, M) = (1 + z_\mathrm{mm})^\beta \mathrm{frac}(M) \Phi_\mathrm{blue}(z_\mathrm{mm}, M).
\end{equation}

Equation \ref{eq:mergers_z} shows that we need to model the redshift dependence of the blue stellar mass function to be able to estimate the major merger contribution to the red sequence. We use the results of \cite{Caplar:2015aa} who derive an analytic expression for the change in $\Phi^{*}_\mathrm{blue}$ and $M^{*}_\mathrm{blue}$ using data from \cite{Ilbert:2013aa}. The redshift evolution of the blue mass function can then be expressed as:

\begin{equation}\label{eq:blue_evolution}
\begin{aligned}
\Phi_\mathrm{blue} (M, z) d\log M = & \ln(10) \Phi^{*}_\mathrm{blue}(z) e^{-(M/M^{*}_\mathrm{blue}(z))} \\
&\times \left (\frac{M}{M^{*}_\mathrm{blue}(z)} \right)^{\alpha_\mathrm{blue} + 1} d\log M\\
\log \Phi^{*}_\mathrm{blue} (z)  = & a_0 + a_1 \kappa + a_2 \kappa^2 + a_3 \kappa^3\\
\log M^{*}_\mathrm{blue}(z)  = & b_0 + b_1 \kappa + b_2 \kappa^2 + b_3 \kappa^3.
\end{aligned}
\end{equation}
with $\kappa = \log(1 + z)$. We adopt the following parameter values from \cite{Caplar:2015aa}: $a_1 = -0.26,\ a_2 = -1.6,\ a_3 = -0.88$ and $b_1 = -0.53,\ b_2 = 3.36,\ b_3 = -3.75$. We adjust $\alpha_\mathrm{blue}$, $a_0$ and $b_0$ so that at $z = 0$ $\Phi_\mathrm{blue}$ corresponds to the best-fit Schechter function of blue galaxies that we determined in \cite{Weigel:2016aa}. We thus assume $\alpha_\mathrm{blue} = \alphablue$, $a_0 = \logphistarblue$ and $b_0 = \logmstarblue$. 

By combining equations \ref{eq:mergers_z} and \ref{eq:blue_evolution} we can model the evolution of $\Phi_\mathrm{red\ mergers}(z, M)$, the stellar mass function of galaxies that have been quenched through major mergers at $z_{\rm mm}$ and are reaching the red sequence at $z$. The number density of merger quenched galaxies that reach the red sequence between $z_1$ and $z_2$ ($z_1 > z_2$) is given by:

\begin{equation}
\begin{aligned}
\Delta \Phi_{\rm red\ mergers}(z_1, z_2, M) =& \Phi_{\rm red\ mergers}(z_1, M) \\
&- \Phi_{\rm red\ mergers}(z_2, M).  
\end{aligned}
\end{equation}
The central panel of Fig. \ref{fig:red_blue_toy_model} shows $\Delta \Phi_{\rm red\ mergers}$ for a range of $z_1$ and $z_2$ values. 

\subsubsection{The evolution of the red mass function}

To estimate the contribution of major merger quenched galaxies to the red sequence within the last $\sim 5\ \rm Gyr$, we also need to model the evolution of the mass function of red galaxies. 

To determine the number density of galaxies that reach the red sequence between $z_1$ and $z_2$ we evolve the blue galaxies along the main sequence. Using equation \ref{eq:blue_evolution}, we generate the mass function of blue galaxies at $z = z_1$. Without quenching, these galaxies will continue to evolve along the main sequence and will gain mass. We use the main sequence equation by \cite{Lilly:2013aa} to estimate this change in stellar mass between $z_1$ and $z_2$ and predict the shape of the blue mass function at $z=z_2$. We measure the mass function of newly quenched objects between $z_1$ and $z_2$ by subtracting the true blue mass function at $z=z_2$ from our prediction. We express the number density of newly quenched objects between $z_1$ and $z_2$ in the following way:

\begin{equation}\label{eq:red_evolution}
\begin{aligned}
M^{*}(z_2) = & M^{*}_\mathrm{blue}(z_1) \cdot \left(1 + sSFR(z_1) \cdot t\right)\\
\Phi_\mathrm{predicted\ blue} (z_2, M) = & \ln (10) \Phi^{*}_\mathrm{blue}(z_1) \\
&\times e^{-M/M^{*}(z_2)} \left(\frac{M}{M^{*}(z_2)} \right)^{\alpha_\mathrm{blue}(z_1) + 1}\\
\Delta \Phi_\mathrm{new\ red} (z_1, z_2, M) = & \Phi_\mathrm{predicted\ blue}(z_2) - \Phi_\mathrm{true\ blue}(z_2)\\
\end{aligned}
\end{equation}

Here, $t$ corresponds to the time between $z_1$ and $z_2$ in Gyr. For the predicted blue mass function, we keep $\Phi^{*}$ and $\alpha$ constant and simply shift $M^{*}$ towards higher stellar masses. We show the number density of newly quenched galaxies to the red sequence in the left panel of Fig. \ref{fig:red_blue_toy_model}. 

\subsubsection{Comparing merger quenched galaxies to all quenched galaxies}

With equations \ref{eq:mergers_z} and \ref{eq:red_evolution} we estimate the contribution of major merger quenched galaxies to the red sequence between $0 \leq z \leq 0.5$. We bin in redshift space and estimate the number density of all quenched and of merger quenched galaxies by computing:
\begin{equation}
\begin{aligned}
\Phi_{\sum \mathrm{new\ red}}(M) =& \sum_i^{N_\mathrm{z-bins}} \Delta \Phi_\mathrm{red}(z_i, z_i - \Delta z, M),\\
\Phi_{\sum \mathrm{red\ mergers}}(M) =& \sum_i^{N_\mathrm{z-bins}} \Delta \Phi_\mathrm{red\ mergers} (z_i, z_i - \Delta z, M). 
\end{aligned}
\end{equation}

The major merger contribution to the build up of red sequence is then given by:
\begin{equation}
\mathrm{red\ merger\ fraction} (M) =\frac{\Phi_{\sum \mathrm{red\ mergers}}(M) }{\Phi_{\sum \mathrm{new\ red}}(M)}.
\end{equation}
We show the red merger fraction as a function of stellar mass in the right-hand panel of Fig. \ref{fig:red_blue_toy_model}. When interpreting the results it is important to be aware that at the high mass end, both the red and the major merger $z \sim 0$ mass functions are affected by significant uncertainties. For the major merger mass function this is also the case at the low mass end (see Fig. \ref{fig:comparison_red_red_ET_mm}). These uncertainties also affect the red merger fraction shown in Fig. \ref{fig:red_blue_toy_model}. At these masses we thus show the red merger fraction with a dashed line. 

As a reference, we show the evolution of the blue mass function and the mass function of merger quenched galaxies reaching the red sequence in Fig. \ref{fig:red_blue_toy_model_app}. With decreasing redshift the number density of blue galaxies increases. Yet even though $\Phi_{\rm{red\ mergers}}$ depends on $\Phi_{\rm{blue}}$, the number density of red merger quenched galaxies decreases with decreasing redshift. This is due to the $(1+z)^{\beta}$ factor which we have introduced in equation \ref{eq:mergers_z}. $(1 + z)^{\beta}$ decreases more steeply than $\Phi^{*}_{\rm{blue}}$ increases. This causes fewer merger quenched galaxies to reach the red sequence at lower redshift. 

Once we have determined $\Phi_{\sum \mathrm{new\ red}}(M)$ we can also estimate the fraction of red galaxies that has been  quenched within the last 5 Gyr. We integrate $\Phi_{\sum \mathrm{new\ red}}(M)$ and the mass function of all red galaxies (see Table \ref{tab:parameters}) between $10^9\ M_\odot$ and $10^{12}\ M_\odot$. Over this mass range, galaxies that have been  quenched within $0 \leq z \leq 0.5$ make up about $41\%$ of all red galaxies. 

We conclude that based on the approach presented here the contribution of major merger quenched galaxies to the build up of the red sequence within the last 5 Gyr is $1 - 5\%$ at a given mass. Within this time range a significant fraction of galaxies has been quenched, the contribution of merger quenched galaxies is however negligible. This is in agreement with our results for the local Universe where we found the number flux of green early type galaxies across the green valley to be low compared to the total green valley flux (see Section \ref{sec:green_valley_flux}).   

\subsection{The significance of major merger quenching}
We discussed the significance of major merger quenching in Sections \ref{sec:compare_red}, \ref{sec:compare_green}, \ref{sec:green_valley_flux} and \ref{sec:red_sequence}. Our first test consisted of comparing the major merger to the red early type mass function. Their different shapes let us to conclude that it is unlikely that all red early type galaxies have been created through through the major merger quenching process that we observe today. Our second analysis was purely based on the stellar mass functions of green galaxies and of major mergers. By comparing their shapes we concluded that the local green valley population is dominated by galaxies that are unlikely to have been quenched through major mergers. This straight forward argument is independent of the mass functions of later stages along the merger sequence and independent of the early type morphological classifications. For our third analysis we used the stellar mass functions of green early types and of green galaxies. We argued that green early types, which are likely to have been merger quenched, only make up a small fraction of all green galaxies. Even if we take into account that early types tend to transition the green valley on shorter time scales than the total green population, merger quenched galaxies are unlikely to make up the majority of the overall green valley flux at $z\sim0$. For our fourth approach we modelled the evolution of the major merger and the red mass function from to $z=0.5$ to $z=0$. We concluded that merger quenched galaxies account for $1-5\%$ of all galaxies that quenched within the last 5 Gyr. 

All four tests are independent from each other, yet their results are consistent: merger quenching is unlikely to contribute significantly to the quenching of galaxies at $z\sim0$ and is unlikely to have quenched the majority of galaxies that reached the red sequence within the last 5 Gyr. To explain the existence of the green valley and the red sequence populations, at least one additional quenching mechanism has to exist. To account for the slow transition rate of green valley galaxies, alternative quenching channels are likely to lead to a change in colour that is slow compared to major merger quenching. As have discussed in Section \ref{sec:compare_red}, alternative quenching processes could also lead to the formation of red early type galaxies, thereby explaining the shape of the red early type mass function.

\section{Discussion}\label{sec:discussion}
\subsection{The role of AGN in the quenching of star formation}
In Section \ref{sec:analysis} we argued that major merger quenching alone is unlikely to account for the quenching of all galaxies that we observe in the local Universe. Alternative quenching channels have to be introduced to explain the properties of the total green valley and red sequence population. For example the slow evolution of green late types is likely to be caused by a quenching process that does not involve major mergers. As we discussed in Section \ref{sec:mass_env_qu}, quenching could also be caused by secular processes \citep{Kormendy:2004aa, Masters:2011aa, Cheung:2013aa}, environmental processes such as ram pressure stripping \citep{Gunn:1972aa} or strangulation \citep{Larson:1980aa, Balogh:2000aa} or AGN feedback \citep{Silk:1998aa, Di-Matteo:2005aa,Schawinski:2006aa, Kauffmann:2007aa, Georgakakis:2008aa, Hickox:2009aa, Cattaneo:2009ab, Fabian:2012aa, Bongiorno:2016aa, Smethurst:2016aa}.

On one side, AGN feedback seems to be necessary to efficiently transform major merger remnants from blue to red early types. For instance, \cite{Springel:2005ad} show that after a gas-rich merger, AGN feedback is necessary to move the merger remnant from the blue cloud to the red sequence. Without AGN feedback, low levels of star formation ensure that the galaxy remains blue (also see e.g. \citealt{Birnboim:2007aa, Khalatyan:2008aa, Hopkins:2008ab}). Similarly, \cite{Sparre:2016aa} also use hydrodynamical simulations to show that without strong AGN feedback the newly formed galaxy can return to be a star forming late type. Using a phenomenological model, \cite{Kaviraj:2011aa} argue that if only star formation feedback is considered, the gas depletion rate is too low to explain the rapid transition of early type galaxies from blue to red (\citealt{Schawinski:2006aa, Schawinski:2014aa}). AGN are thus often found in morphologically disturbed galaxies or ULIRGs (e.g. \citealt{Urrutia:2008aa, Bennert:2008aa, Yuan:2010aa,  Koss:2011aa, Hong:2015aa}).

On the other side, a major merger does not seem to be necessary for a black hole to be actively accreting \citep{Ellison:2011aa}. For example,  \cite{Treister:2012ab} argue that only the most luminous AGN are triggered through major galaxy mergers.  \cite{Simmons:2013aa} find AGN in massive galaxies without classical bulges. These galaxies are unlikely to have experienced significant mergers in the past which suggests that secular processes might be feeding the central black hole (see e.g. \citealt{Jogee:2006aa, Alexander:2012aa}).   

AGN feedback could thus be involved in two separate quenching channels. First, in major mergers AGN feedback might be a necessary, but not sufficient condition for quenching. Second, if efficient enough, AGN feedback might be sufficient for the quenching of non-merger galaxies.

\subsection{Close pairs}
\citetalias{Darg:2010aa} argue that their major merger catalog probes the post-close pair stage. Introducing a volume limit, they compare a catalog of SDSS close pairs (projected separation $< 30$\ kpc, line-of-sight velocity difference $< 500\ \rm km\ s^{-1}$) to their sample of visually classified mergers. After eliminating pairs which are well separated and show no signs of interaction, $\sim 64\%$ of all perturbed, remaining pairs have $f_{\rm m}< 0.4$ and are thus not part of the \citetalias{Darg:2010aa} sample. Yet \citetalias{Darg:2010aa} also claim a strong correlation between the merger vote fraction and the projected separation of two merging galaxies: the further apart two galaxies are, the less likely they are to be classified as a merger. They thus argue that their selection technique is sensitive to merging systems which show clearer signs of interaction than typical close pair galaxies, therefore probing merging systems which have progressed from the close pair stage.

To compare our results to observations of galaxies in a stage which is likely to proceed the \citetalias{Darg:2010aa} major merger stage, we use the close pair study by \cite{Domingue:2009aa}. Using SDSS (DR5; \citealt{Adelman-McCarthy:2007aa}) and Two Micron All Sky Survey (2MASS, Extended Source Catalog; \citealt{Jarrett:2000aa}) data, \cite{Domingue:2009aa} determine the luminosity function of close pairs. For all pairs, isolated pairs and grouped pairs they find slope $\alpha$ values of $-1.03\pm0.09$, $-0.8\pm0.1$ and $-1.2\pm0.2$, respectively.

With \alphamergers\ we find an increasing slope for major mergers, whereas the luminosity function of all close pairs is almost flat. Our stellar mass function best-fitting parameters are based on the STY results. A shallower mass function for major mergers is thus less likely, yet not completely ruled out (see Fig. \ref{fig:sequence}). The slope of the isolated pair luminosity function is consistent with the slope of our post merger, blue early type and mergers in underdense regions stellar mass functions (see Table \ref{tab:parameters}). The slope of the grouped pair luminosity function is steeper and consistent with the slope of the blue galaxies stellar mass function ($\alpha=-1.21\pm0.01$). 

While our results are not directly comparable to the work by \cite{Domingue:2009aa}, it is interesting to note that the stellar mass and luminosity functions of close pairs and galaxies along the merger sequence are unusually flat. If close pairs and major mergers were randomly drawn from the entire or the blue galaxy population, we would expect their luminosity and mass function to have comparable, steeper slopes.This difference is slopes is also reflected in the mass dependent merger fractions (see Fig. \ref{fig:fraction}). As we discussed in  Section \ref{sec:merger_frac}, the unusually shallow mass functions of close pairs and subsequent stages could be caused by a mass independent dark matter halo merger fraction in combination with a mass dependent stellar-to-halo mass conversion \citep{Bertone:2009aa, Hopkins:2010aa,Hopkins:2010ab, Behroozi:2013aa}.

\subsection{Slow and fast rotators}
We used the colour of galaxies in combination with their morphological classifications to study the process of merger quenching. Morphological classifications were essential for our approach as the allowed us to follow galaxies that are likely to have been merger quenched along their evolution from the blue cloud to the red sequence. However we have not yet included kinematic information in our approach. After projects such as SAURON \citep{de-Zeeuw:2002aa}, CALIFA \citep{Sanchez:2012aa} and $\rm{ATLAS}^{\rm{3D}}$ \citep{Cappellari:2011ab}, the next generation of IFU surveys, for example the SAMI galaxy survey \citep{Croom:2012aa, Bryant:2015aa}, MaNGA \citep{Bundy:2015aa} and the Hector survey \citep{Bland-Hawthorn:2015aa}, will increase the number of observed galaxies by orders of magnitude. These surveys will allow us to move from studying single objects to including kinematic information in a statistical approach. In the context of quenching the formation and evolution of slow and fast rotating galaxies (\citealt{Illingworth:1977aa, Davies:1983aa} and e.g. \citealt{Cappellari:2016aa} and references therein) will be especially important. The formation of fast and slow rotators has been studied observationally (e.g. \citealt{Oh:2016aa, Forbes:2016aa}) and with numerical (e.g. \citealt{Bois:2010aa, Bois:2011aa}), semi-analytic (e.g. \citealt{Khochfar:2011aa}) and hydrodynamical simulations (e.g. \citealt{Naab:2014aa}). While there seems to be a clear link between major mergers and slow and fast rotators, no simple model regarding their formation has emerged so far. Among other parameters, the amount of involved gas, the number of mergers in the past, the mass ratio, the amount and orientation of angular momentum of the merging galaxies and the resulting spin up or spin down of the remnant seem to play a significant role in their evolution. Nonetheless, similar to morphologies, considering  kinematics in the analysis of quenching will be very insightful. We will thus be exploring the role of slow and fast rotators in the context of merger quenching in future work.

\section{Summary}\label{sec:summary}
We used SDSS DR7 and Galaxy Zoo 1 data in combination with visually selected major merger and post merger samples to determine the stellar mass functions of major mergers and post mergers in the local Universe (see Fig. \ref{fig:merger_post_merger_smf}). These mass functions allowed us to constrain the fraction of major merger and post mergers relative to the entire galaxy sample and to blue and to red galaxies (see Fig. \ref{fig:fraction}). In Section \ref{sec:mass_env_qu} we compared our measurement of the major merger mass function to the empirical quenching model by \citetalias{Peng:2010aa}. We concluded that the major merger mass function in the local Universe in inconsistent with the mass independent, but environment dependent merger quenching process that \citetalias{Peng:2010aa} propose.

To investigate the process of major merger quenching and its significance in the local Universe, we made three key assumptions (see Section \ref{sec:method}):
\begin{enumerate}
	\item {merger quenched galaxies pass through five distinct stages: major merger, post merger, blue early type, green early type and red early type,}
	\item {the probability of a galaxy evolving from the major merger to the red early type stage is mass independent and effects such as disc reforming, AGN feedback, dynamical friction and the fading of merger features do not introduce a mass dependent bias,}
	\item {while transitioning from the major merger to the red early type stage galaxies do not gain significant amounts of stellar mass.}
\end{enumerate}
These three assumptions led us the to the expectation that the stellar mass functions of galaxies in these five stages should be similar in shape if they represent an evolutionary sequence.

In addition to the major merger and post merger samples, we also determined the mass functions for galaxies in the blue, green and red early type stage. Using the ratio of mass functions of subsequent stages, we showed that galaxies in the major merger, post merger, and blue and green early type stage have indeed similar mass function shapes (see Fig. \ref{fig:trans_curves}). We concluded that that major mergers are likely to lead to the quenching of star formation with the five stages that we considered representing an evolutionary sequence from star formation to quiescence. The flat transition curves furthermore provide support for our assumption that the probability of merger remnants reaching the red early type stage is likely to be mass independent.

The mass functions of galaxies along the sequence allowed us to constrain the relative amount of time spent in the major merger, post merger, blue early type and green early type stage (see Fig. \ref{fig:timescales}). We assumed that the majority of galaxies transition from one stage to the next. Based on this assumption we found that galaxies spend $\sim 60\%$ of their time in the green early type and $\sim 5\%$ of their time in the post merger stage.

To investigate the significance of major merger quenching in the local Universe we used four tests which vary in their level of sophistication and assumptions: 
\begin{enumerate}
	\item We compared the shapes of the major merger and the red early type mass function (see Fig. \ref{fig:comparison_red_red_ET_mm}) and concluded that it is unlikely that all red early types were created through the major merger quenching process that we observe today. Alternative quenching channels are necessary to explain the red early type mass function shape.
	\item By comparing major merger and the green mass function shapes (see Fig. \ref{fig:comp_mm_green_mf}) we argued that the green valley population is dominated by galaxies which are unlikely to have been major merger quenched in the past. 
	\item We estimated the contribution of major merger quenched galaxies to the overall green valley number flux (see Fig. \ref{fig:green_valley_frac}). We concluded that green early type galaxies, which are likely to have been major merger quenched, are unlikely to dominate the $z\sim0$ flux of galaxies across the green valley.
	\item For our final test we simulated the evolution ot the red stellar mass function. We estimated the fraction of galaxies that are likely to have been merger quenched within the last 5 Gyr (see Fig. \ref{fig:red_blue_toy_model}) to be $1-5\%$ at a given stellar mass. 
\end{enumerate} 

In summary, our analysis shows that major mergers are likely to lead to an evolution from star formation to quiescent via quenching. Yet merger quenching is unlikely to account for the majority of  quenching, neither at $z\sim0$ nor within the last 5 Gyr. To explain the existence of the green valley and red sequence population alternative quenching channels, which are likely to lead to a slow green valley transition, have to exist.

\bibliographystyle{apj}
\bibliography{Bib_Lib.bib}
We thank the anonymous referee for helpful comments. AKW and KS gratefully acknowledge support from Swiss National Science Foundation Grants PP00P2\_138979 and PP00P2\_166159. BDS acknowledges support from the National Aeronautics and Space Administration (NASA) through Einstein Postdoctoral Fellowship Award Number PF5-160143 issued by the Chandra X-ray Observatory Center, which is operated by the Smithsonian Astrophysical Observatory for and on behalf of NASA under contract NAS8-03060. This research made use of NASA's ADS Service. This publication made use of Astropy, a community-developed core \textsc{Python} package for Astronomy \citep{2013A&A...558A..33A} and the \textsc{Python} plotting package matplotlib \citep{Hunter:2007}. This publication made extensive use of the Tool for OPerations on Catalogues And Tables (\textsc{TOPCAT}), which can be found at \url{http://www.starlink.ac.uk/topcat/}. The development of Galaxy Zoo was supported by the Alfred P. Sloan Foundation and The Leverhulme Trust. Funding for the SDSS has been provided by the Alfred P. Sloan Foundation, the Participating Institutions, the National Aeronautics and Space Administration, the National Science Foundation, the US Department of Energy, the Japanese Monbukagakusho, and the Max Planck Society. The SDSS website is http://www.sdss.org/. The SDSS is managed by the Astrophysical Research Consortium (ARC) for the Participating Institutions. The Participating Institutions are The University of Chicago, Fermilab, the Institute for Advanced Study, the Japan Participation Group, The Johns Hopkins University, Los Alamos National Laboratory, the Max-Planck-Institute for Astronomy (MPIA), the Max-Planck-Institute for Astrophysics (MPA), New Mexico State University, University of Pittsburgh, Princeton University, the United States Naval Observatory and the University of Washington.

\appendix
\section{Extended analysis}

\subsection{The effect of stellar mass estimates on the shape of the major merger mass function}
\label{sec:merger_smf_mass}

\begin{figure*}
	\includegraphics[width=\textwidth]{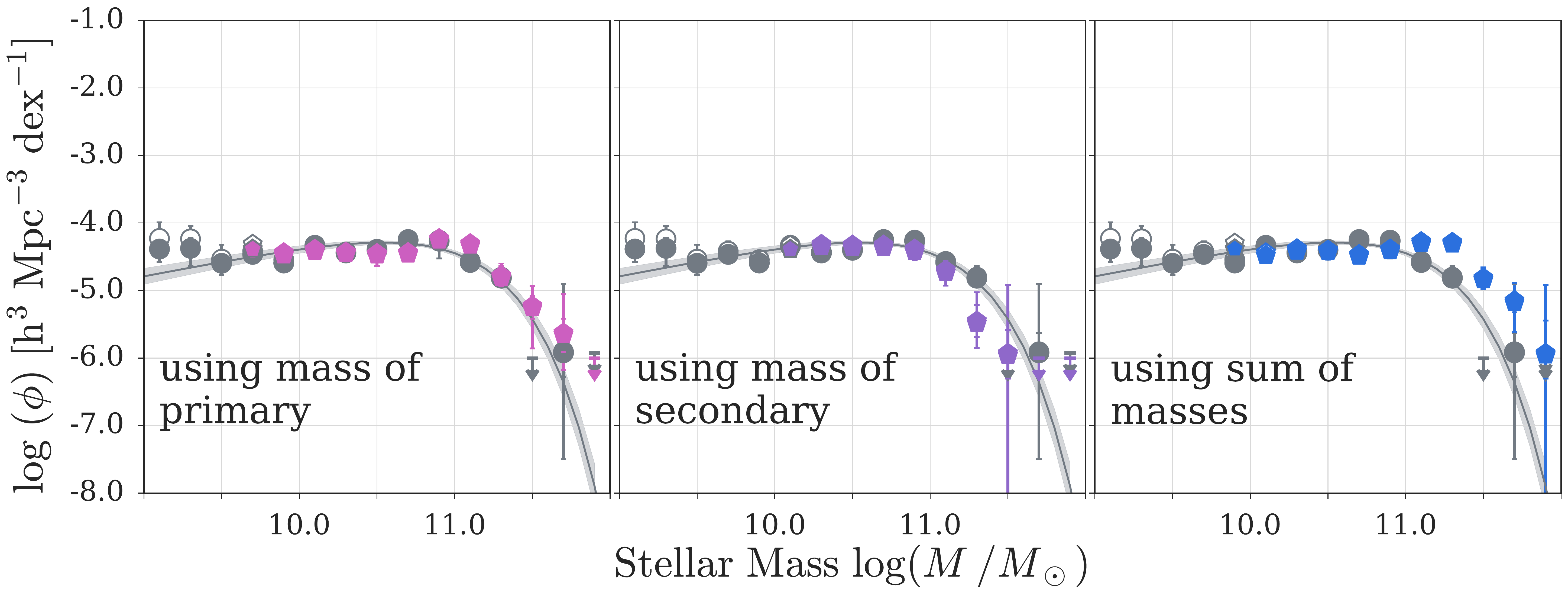}
	\caption{\label{fig:merger_smf_mass}The effect of using the masses of the primaries, the secondaries and the sum  of the masses of the merging galaxies for the major mass function construction. To determine the major merger mass function which we use for our analysis, which we show in grey here, we use mass estimates from the MPA JHU catalog  \protect\citep{Kauffmann:2003ab, Brinchmann:2004aa, Salim:2007aa}. As spectra are only available for some of the galaxies in the \protect\citetalias{Darg:2010aa} catalog, this results in us using a mix of primary and secondary masses for the mass function determination. We use the photometry based stellar mass estimates by \protect\citetalias{Darg:2010aa} to show that this combination of mass estimates does not significantly affect the major merger mass function shape. In the left-hand panel we show the major merger mass function based on the mass of the more massive merging partner. For the stellar mass function in the central panel we used the mass of the secondaries as input. In the right-hand panel we show the stellar mass function based on the sum of masses.}
\end{figure*}

\begin{figure*}
	\includegraphics[width=\textwidth]{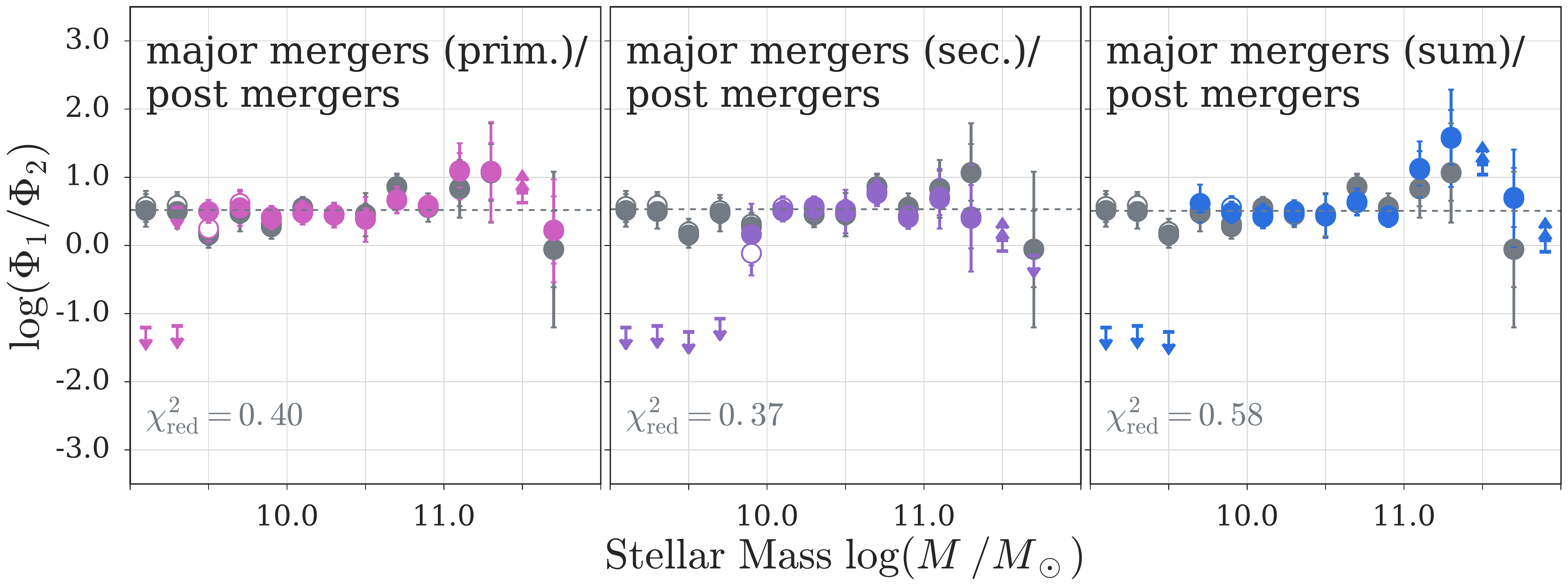}
	\caption{\label{fig:trans_mass}The ratio between the modified major merger mass functions and the post merger mass function. We use the major merger mass functions shown in Fig. \ref{fig:merger_smf_mass} to construct the transition curves (see Section \ref{sec:mm_mf}) relative to the post merger sample. From left to right and in analogy to Fig. \ref{fig:merger_smf_mass}, we use the mass function based on primary masses, secondary masses and the sum of masses for the ratio. The horizontal dashed line shows the best fitting relation if we assume a constant ratio between these modified merger mass functions and the post merger mass function. The corresponding $\chi^{2}_{\rm red}$ values are given within the panels. Shown in grey is the ratio between the major merger mass function which we use for our analysis and the post merger mass function. Even though using the sum of stellar masses as input for the merger mass function leads to a shift in $M^{*}$ (see right-hand panel of Fig.  \ref{fig:merger_smf_mass}), the transition curve for mergers and post-mergers shows no significant mass dependence, as the right-hand panel shows.}
\end{figure*}

As we discussed in Section \ref{sec:merger_sample}, spectra for both merging galaxies are only available for $23\%$ of all merging systems in the \citetalias{Darg:2010aa} catalog. \citetalias{Darg:2010aa} thus fit two-component star formation histories to the photometry to estimate the stellar masses of all merging galaxies. We use these stellar mass estimates to select major mergers. However, for the construction of the major merger mass function, which is for instance shown in Fig.  \ref{fig:sequence}, we use stellar mass estimates from the MPA JHU catalog \citep{Kauffmann:2003ab, Brinchmann:2004aa, Salim:2007aa} to ensure consistency with the other mass functions used for our analysis. If both galaxies that are involved in a major merger have been observed spectroscopically, we use the mass of the more massive galaxy, the primary for the mass function construction. If a spectrum is only available for one of the two merging galaxies, we use the mass of the galaxy for which a spectrum is available. According to the \citetalias{Darg:2010aa} mass measurements, in $69\%$ of all cases the galaxy with the available spectrum corresponds to the primary. We now illustrate the effect of this approach on the shape of the major merger mass function.

In Fig. \ref{fig:merger_smf_mass} we show three versions of the major merger mass function. Each is based on a different set of stellar mass function estimates. For the stellar mass function in the left-hand panel we used the \citetalias{Darg:2010aa} stellar mass values of the primaries as input. In the central panel we show the merger mass function based on the \citetalias{Darg:2010aa} mass estimates for the less massive of the two merging galaxies, the secondary. For the stellar mass function in the right-hand panel we used the sum of the primary and the secondary as input. As spectra are only available for a subset of galaxies in the \citetalias{Darg:2010aa} sample, we use photometric redshifts if spectroscopic redshifts are unavailable. For the stellar mass function shown in the right-hand panel we use the redshift and the apparent magnitude of the primary to determine if the merging system lies within our redshift range and to correct for mass completeness effects \citep{Weigel:2016aa}. In Fig. \ref{fig:merger_smf_mass} we show the major merger mass function that was estimated with the approach discussed above in grey for comparison. This is the mass function that we use for our analysis.

Fig. \ref{fig:merger_smf_mass} shows that the merger mass functions that are solely based on the masses of the primaries and secondaries do not deviate significantly from the stellar mass function that we use for our analysis which is generated with a mix of primary and secondary masses.

As expected, using the sum of the merging galaxies for the stellar mass function construction leads to a deviation at the steep high mass end. To test if this shift in $M^{*}$ significantly affects our analysis, we determine the ratio between the major merger mass function and the post merger mass function (see Section \ref{sec:mm_mf}). From left to right, Fig. \ref{fig:trans_mass} shows the major merger mass function based on primary masses, secondary masses and the sum of primary and secondary masses relative to the post merger mass function. Shown in grey is the ratio between the major merger mass function which we use for our analysis and the post merger mass function. The horizontal dashed lines illustrate the best fitting relation if we assume a constant ratio between the modified merger mass functions and the post merger mass function. The corresponding $\chi^{2}_{\rm reduced}$ are shown within in the panels.

Fig. \ref{fig:trans_mass} shows that even though using the sum of masses for the major merger mass function construction leads to a shift in $M^{*}$, the ratio between the major merger and the post merger mass function still shows no significant mass dependence. Using the sum of masses for the merger mass function would thus result in a steeper mass dependence of the merger fraction (see Fig. \ref{fig:fraction}), but it would not significantly impact our analysis regarding major merger quenching.

\subsection{Merger contribution to the build up of the red sequence}

\begin{figure*}
	\includegraphics[width=\textwidth]{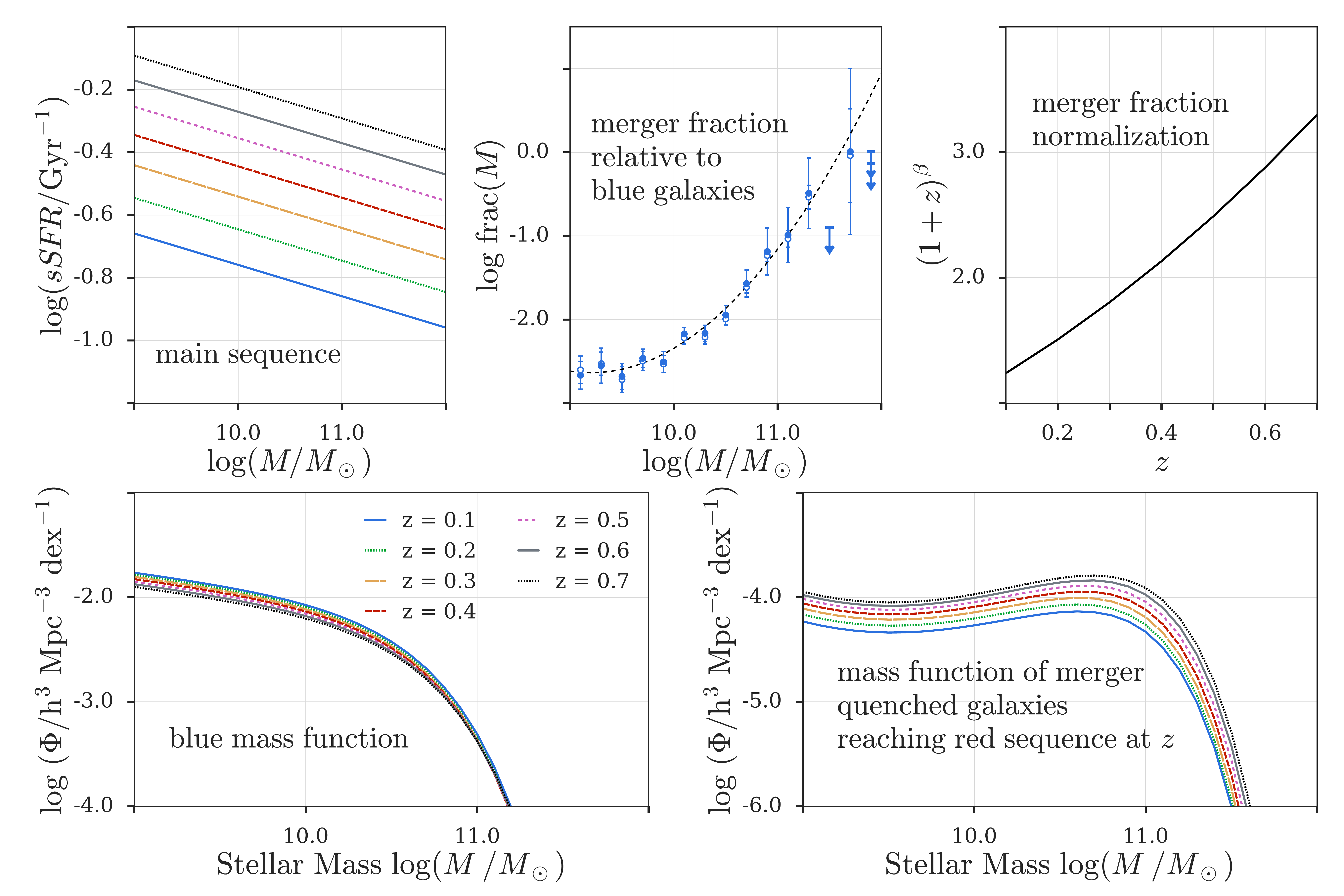}
	\caption{\label{fig:red_blue_toy_model_app} Input for the simple model to estimate the contribution of merger quenched galaxies to the build up of the red sequence within $0 < z < 0.5$ (see Section \ref{sec:red_sequence}). We base our model on the redshift evolution of the blue stellar mass function (bottom left-hand panel, \protect\citealt{Caplar:2015aa}). To estimate the number density of newly quenched galaxies between $z_1$ and $z_2$ ($z_1 > z_2$), we evolve $M^{*}(z_1)$ along the main sequence (top right-hand panel, \protect\citealt{Lilly:2013aa}) and subtract the true, observed blue mass function at $z_2$ from the predicted one. We assume that it takes merger quenched galaxies $\sim 1\ \rm Gyr$ to transition form the major merger to the red early type stage. To determine the evolution of these red merger quenched galaxies, we parametrize the major merger mass function as the product of the blue mass function and the merger fraction  at $z\sim 0$ (central top panel). Furthermore, we ensure that the number of major mergers increases with increasing redshift by introducing a factor of $(1 + z)^{\beta}$ ($\beta = 2.25$, top right-hand panel, \citealt{Bridge:2010aa}). The redshift evolution of merger quenched galaxies mass function is shown in the bottom right-hand panel. Note that for the red major merger mass function at $z$ we use $\Phi_{\rm blue}(z_{\rm mm})$ and $(1 + z_{\rm mm})$ where $z_{\rm mm} - z  \mathrel{\hat{=}} 1$\ Gyr. We thus show $\Phi_{\rm blue}$, the main sequence and $(1 + z)^{\beta}$ for values beyond $z = 0.5$.}
\end{figure*}

\begin{figure}
	\includegraphics[width=\columnwidth]{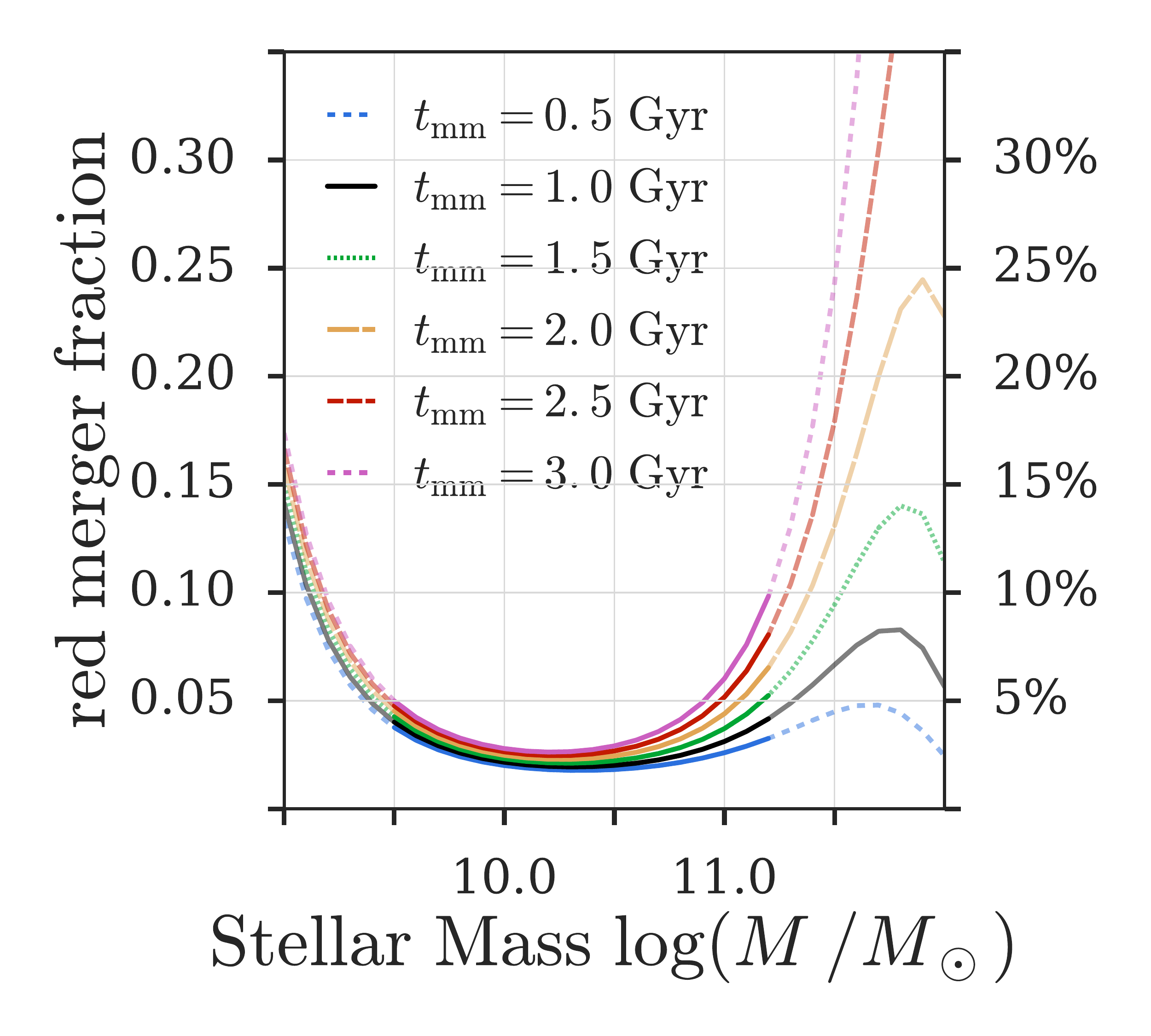}
	\caption{\label{fig:diff_timescale}Contribution of major merger quenched galaxies to the build up of the red sequence within the last 5 Gyr. In analogy to Fig. \ref{fig:red_blue_toy_model}, we show the the fraction of red galaxies that have been quenched through major mergers as a function of stellar mass. For our model in Sec. \ref{sec:red_sequence} in we assume that it takes galaxies of the order of $t_{\rm mm} \sim 1\ \rm Gyr$ to transition from the major merger to the red early type stage.  Here we show that our result is not significantly affected by the chosen $t_{\rm mm }$ value. Note that the low and the high mass end of the major merger mass function is affected by significant uncertainties. We thus only consider stellar masses within $9.5 < \log(M/M_\odot) < 11.25$ for our model.}
\end{figure}

In Section \ref{sec:red_sequence}  we determine the contribution of major merger quenched galaxies to the build up of the red sequence within the last 5 Gyr. To do so, we model the redshift evolution of the major merger mass function and estimate the number of newly quenched galaxies within a given redshift interval. In addition to Fig. \ref{fig:red_blue_toy_model}, we show the evolution of the main sequence, the merger fraction as a function of stellar mass, the redshift evolution of the normalization of the merger fraction and the evolution of the blue mass function in Fig. \ref{fig:red_blue_toy_model_app}. Furthermore, we show the evolution of the mass function of galaxies that have been quenched through major mergers and reach the red sequence at a certain redshift. 

For our simple model, we assume that it takes galaxies of the order of $t_{\rm mm} \sim 1\ \rm Gyr$ to evolve from the major merger to the red early type stage. In Fig. \ref{fig:diff_timescale} we show that our estimate of the contribution of major merger quenched galaxies to the red sequence since $z \sim 0.5$ is not significantly affected by the chosen $t_{\rm mm}$ value.

\end{document}